\documentclass[10pt,journal,compsoc]{IEEEtran}
\ifCLASSOPTIONcompsoc
  \usepackage[nocompress]{cite}
\else
  \usepackage{cite}
\fi
\ifCLASSINFOpdf
\else
\fi
\usepackage[cmex10]{amsmath}
\ifCLASSOPTIONcompsoc
 \usepackage[caption=false,font=footnotesize,labelfont=sf,textfont=sf]{subfig}
\else
 \usepackage[caption=false,font=footnotesize]{subfig}
\fi
\usepackage[hyphens]{url}

\usepackage{multirow}
\usepackage{makecell}
\usepackage{color}
\usepackage{graphicx}
\graphicspath{ {images/} }

\usepackage{amsthm}
\theoremstyle{definition}

\usepackage[normalem]{ulem}
\usepackage{tikz}
\usepackage{textcomp}
\usepackage{hyperref}
\usepackage{lipsum}

\usepackage{soul}

\newcommand{\DEFAULTHIGHLIGHT}[1]{\textcolor{red}{#1}}

\newcommand\copyrighttext{%
  \footnotesize \textcopyright 2019 IEEE. Personal use of this material is permitted. Permission from IEEE must be obtained for all other uses, in any current or future media, including reprinting/republishing this material for advertising or promotional purposes, creating new collective works, for resale or redistribution to servers or lists, or reuse of any copyrighted component of this work in other works. DOI: \href{https://doi.org/10.1109/TSC.2019.2928551}{10.1109/TSC.2019.2928551}}
\newcommand\copyrightnotice{%
\begin{tikzpicture}[remember picture,overlay]
\node[anchor=south,yshift=10pt] at (current page.south) {\fbox{\parbox{\dimexpr\textwidth-\fboxsep-\fboxrule\relax}{\copyrighttext}}};
\end{tikzpicture}%
}

\begin{document}

%
\title{A Comparative Measurement Study of \\ Deep Learning as a Service Framework}
%
%
%
%


\author{\small{Yanzhao~Wu,
        Ling~Liu,~\IEEEmembership{Fellow,~IEEE,}
        Calton~Pu,~\IEEEmembership{Fellow,~IEEE,}
        Wenqi~Cao, Semih~Sahin, Wenqi~Wei, Qi~Zhang}
\IEEEcompsocitemizethanks{\IEEEcompsocthanksitem Y. Wu, L. Liu, C. Pu, W. Cao, S. Sahin and W. Wei are with the School of Computer Science, Georgia Institute of Technology, Atlanta, GA, 30332.
E-mail: \{lingliu, calton.pu\}@cc.gatech.edu, \{yanzhaowu, wcao39, ssahin7, wenqiwei\}@gatech.edu
\IEEEcompsocthanksitem Q. Zhang is with IBM T. J. Watson, New York, USA. E-mail: Q.Zhang@ibm.com}
\thanks{Manuscript received X; revised Y.}}

\IEEEtitleabstractindextext{%
\begin{abstract}
Big data powered Deep Learning (DL) and its applications have blossomed in recent years, fueled by three technological trends: a large amount of digitized data openly accessible, a growing number of DL software frameworks in open source and commercial markets, and a selection of affordable parallel computing hardware devices. However, no single DL framework, to date, dominates in terms of performance and accuracy even for baseline classification tasks on standard datasets, making the selection of a DL framework an overwhelming task.
This paper takes a holistic approach to conduct empirical comparison and analysis of four representative DL frameworks with three unique contributions.
{\em First}, given a selection of CPU-GPU configurations, we show that for a specific DL framework, different configurations of its hyper-parameters may have a significant impact on both performance and accuracy of DL applications.
{\em Second}, to the best of our knowledge, this study is the first to identify the opportunities for improving the training time performance and the accuracy of DL frameworks by configuring parallel computing libraries and tuning individual and multiple hyper-parameters.
{\em Third}, we also conduct a comparative measurement study on the resource consumption patterns of four DL frameworks and their performance and accuracy implications, including CPU and memory usage, and their correlations to varying settings of hyper-parameters under different configuration combinations of hardware, parallel computing libraries. 
We argue that this measurement study provides in-depth empirical comparison and analysis of four representative DL frameworks, and offers practical guidance for service providers to deploying and delivering DL as a Service (DLaaS) and for application developers and DLaaS consumers to select the right DL frameworks for the right DL workloads.
\end{abstract}

\begin{IEEEkeywords}
Deep Learning as a Service; Big Data; Deep Neural Networks; Accuracy
\end{IEEEkeywords}}

\maketitle
\copyrightnotice

\IEEEdisplaynontitleabstractindextext

%
\IEEEpeerreviewmaketitle

\section{Introduction}\label{section:introduction}

%
%
%
%

\IEEEPARstart{B}{ig} data powered Deep Learning (DL) systems and applications are gaining huge popularity recently, in numerous fields, such as image classification, object detection, machine translation and NLP.
We witness two emerging trends: (1) a growing number of DL software frameworks in both open source and commercial markets, represented by Tensor-Flow~\cite{tensorflow}, Caffe \cite{caffe}, Torch \cite{torch}, Theano \cite{theano}, CNTK \cite{cntk}, Keras \cite{keras} and PyTorch \cite{pytorch}
and (2) an increasing volume and variety of DL applications with diverse datasets and domain-specific DL problems. It is widely recognized that choosing the right DL framework for the right applications becomes a daunting task for many researchers, developers and domain scientists. Although there are some existing DL benchmarking efforts, most of them have centered on studying different CPU-GPU configurations and their impact on different DL frameworks with standard datasets~\cite{awan-caffe-variants, dawnbench, shi-benchmarking-state-of-the-art, shams-evaluation-HPC}. Even under the same CPU-GPU configuration, no single DL framework dominates the performance and accuracy for standard datasets, such as MNIST~\cite{mnistlenet}, CIFAR~\cite{cifar10-100}, ImageNet~\cite{imagenet-2015}. Little efforts have been engaged to systematically study the impacts and correlations of various hardware configurations, parallel computing libraries, and DL hyper-parameters on both the performance and the accuracy of DL frameworks for different datasets and DL applications, and how system resources, e.g., CPU and memory, are consumed and contribute to the performance and accuracy of DL frameworks and applications. 

Bearing these problems in mind, in this paper we take a holistic approach to design and conduct a comparative measurement study of four representative DL frameworks, focusing on how they optimize their performance and accuracy using popular DL datasets and workloads. This paper makes three original contributions. 
{\em First}, our empirical study shows that careful selection of hardware configurations, parallel computing libraries, and \textit{hyper-parameters} can have a significant impact on both the performance and the accuracy of DL frameworks for any given DL workloads under a fixed CPU-GPU hardware configuration. The optimal configuration of hyper-parameters is highly dependent on a number of parameters, such as the choice of specific DL framework, the specifics of the datasets and the learning tasks, the structural complexity of deep neural networks, and their specific parallel computation implementation libraries. Thus, the optimal settings of hyper-parameters for one DL framework (e.g., TensorFlow) often do not work well for another DL framework (e.g., Caffe or Torch), or for different datasets or different learning tasks under the same DL framework with the dual goals of high performance and high accuracy.
{\em Second}, to the best of our knowledge, this study is the first to identify the opportunities for configuring parallel computing libraries and tuning individual and multiple hyper-parameters for improving the training time performance and the accuracy of DL frameworks. Through systematic experiments, our study shows that the runtime performance and accuracy of DL frameworks can be significantly improved.
{\em Third}, we analyze the resource consumption patterns, such as CPU and memory usage, under different configuration combinations of hardware, parallel computing libraries, and hyper-parameters. We show the resource consumption implications for different hardware configurations, different parallel computing libraries, and 
different configurations of hyper-parameters, including the default configurations used by existing open source DL frameworks. 

For example, we show that the mini-batch (batch) size and the number of iterations (\#Iterations or \#Epochs) are the two hyper-parameters that have significant impact on both performance and accuracy of the four DL frameworks in our study. Furthermore, learning rate can have significant impact on the accuracy of the DL frameworks for many datasets and learning tasks. 
Although larger batch sizes and a larger number of iterations directly correlate to the training time for all DL frameworks, their impact on accuracy varies for different DL frameworks and often exhibits non-linear relationship. Furthermore, larger number of GPUs and higher capacity of memory and CPU may not result in shorter training time and better accuracy.

The comparative analysis and insights obtained from our in-depth empirical measurement study provides three unique benefits to the big-data services and DL as a service (DLaaS) communities: (1) It provides evidences and recommendations for DL as a Service developers to further enhance the parallel computation libraries and the programmable capabilities of DL frameworks to better leverage the advancement of GPU hardware capabilities and capacities; (2) It offers practical guidelines for service providers to effectively deploying and delivering DLaaS to their diversed applications and consumers; and (3) It steers DLaaS users and application developers to select the right DL frameworks for the right DL workloads.

\section{Reference Model for DL Frameworks}
All DL frameworks encapsulate a chosen deep neural network (DNN) to learn and generate a DNN model over the training dataset $D$. Thus, a DL framework can be abstracted as a function $y=f_\theta(x)$, where $\theta$ denotes the model parameters, $x \in \mathcal{R}^{n}$ represents an $n$-dimension input and $y\in \mathcal{R}^{m}$ denotes the output, an $m$-dimension vector. The DNN typically contains a large number of model parameters ($\theta$), such as the neuron weights and hyper-parameters. Hyper-parameters primarily include the batch size, the number of training iterations and the learning rate (LR). For example, Inception-v3 contains almost 25 million model parameters~\cite{inception-v3}. Thus, in practice, the $\theta$ is first initialized with a set of values (random or fixed), then tuned by the training phase. 

Mathematically, for a training dataset $D$, the objective of training is to minimize the average loss over all $|D|$ \cite{caffe} in each iteration:
 \vspace{-4mm}
 \begin{equation}
 \small
 L(W) = \frac{1}{|D|} \sum_i^{|D|} l_W\left(X_i\right)
 \label{formula:average-loss}
 \vspace{-3mm}
 \end{equation}
where $W$ represents the neuron weights, $L(W)$ indicates the average loss over all $|D|$, and $l_W$ is the loss on a data sample $X_i$ (computed by the feed-forward process). Since $|D|$ can be very large, in practice, a stochastic approximation of this objective is to use mini-batch (batch). By using $N (\ll |D|)$ samples, the average loss can be estimated as:
 \vspace{-3mm}
 \begin{equation}
 \small
 L(W) \approx \frac{1}{N} \sum_i^{N} l_W\left(X_i\right)
 \label{formula:average-loss-batch}
 \vspace{-3mm}
 \end{equation}
where $N$ denotes the batch size for this iteration. In particular, a typical way to select these $N$ samples is to put $N$ continuous samples from the shuffled $D$ into this batch. After $\frac{|D|}{N}$ iterations, the $D$ is traversed, forming an \textit{epoch}. For each epoch, no overlapping between batches exists and $D$ needs to be reshuffled.

It is further specified by an optimizer, e.g., stochastic gradient descent (SGD) \cite{sgd} and Adam \cite{adam} on how to update DNN parameters $W$. For examples, SGD typically updates $W$ as:
 \vspace{-3mm}
 \begin{equation}
 \Delta W = - \alpha \nabla L(W)
 \vspace{-2mm}
 \end{equation}
where $\alpha$ represents the learning rate, controlling the extent of each update.

\section{DL Frameworks}
\begin{figure}[h!]
  \centering
  \includegraphics[width=0.8\linewidth]{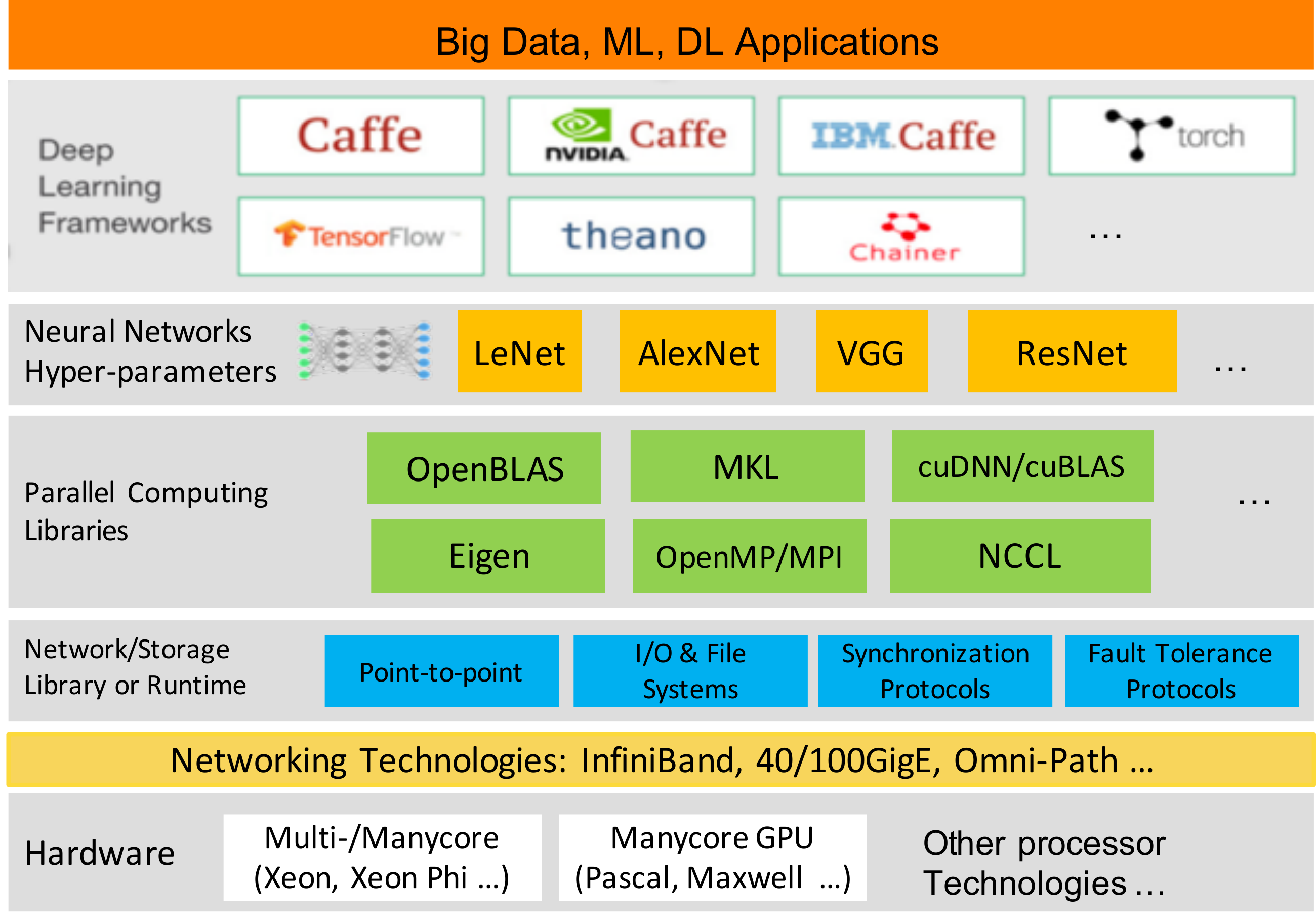}
  \caption{Architectural Overview of the DL Frameworks}
  \label{fig:overview-DL-architecture}
\vspace{-4mm}
\end{figure}

Most DL frameworks, such as TensorFlow, Caffe, Torch, Theano, CNTK, Keras and PyTorch, adopt a similar layered software architecture and provide APIs to enable users to configure the DNN model and the training methods (optimizers). Figure \ref{fig:overview-DL-architecture} shows an overview of the reference architecture used by most of DL frameworks. Most of the existing DL frameworks are implemented on top of popular parallel computing libraries: BLAS (Basic Linear Algebra Subprograms) libraries, such as OpenBLAS~\cite{openblas}, MKL~\cite{mkl}, and cuBLAS~\cite{nvidia}, NCCL (NVIDIA Collective Communications Library)~\cite{nvidia}, OpenMP/MPI~\cite{openmp, mpi} and Eigen~\cite{eigen}. Network and storage tier is on top of the bare metal hardware and connecting to the parallel computation libraries. LeNet, AlexNet, VGG, Resnet are some popular neural network (NN) models offered as user-configurable NN options by most DL frameworks. For example, Caffe, Torch, Theano all provide options of AlexNet, LeNet, VGG, and ResNet. The Google Inception network~\cite{inception-v3} is an extension of LeNet.



\vspace{-3mm}

\cite{dl-framework-ranking} evaluated the popularity of deep learning frameworks from 11 data sources across 7 distinct categories, including Google Search volume, arXiv articles, and GitHub activity. The results show that TensorFlow, Torch (PyTorch), Caffe and Theano are among the Top 5 most popular deep learning frameworks. Moreover, these 4 deep learning frameworks adopt different design principles and delegate state-of-the-art implementation and performance, making them the ideal choice for our study.

Specifically, \textbf{TensorFlow} is developed by Google. It adopts the dataflow graph model. Neurons are modeled as tensors (multi-dimension arrays) flow between nodes via edges. Each node represents a mathematical operation while each edge indicates dataflow. The dataflow model with tensors equips TensorFlow with an ideal API for implementing neural networks. Remarkable flexibility, portability and high efficiency is embedded in its design.
\textbf{Caffe} is developed by BVLC (Berkeley Vision and Learning Center). It provides the model definition in a layer granularity to simplify DL. The network and the training method (solver) are defined in \textit{.prototxt} files separately, which can be used in command lines directly by the independent Caffe program. Caffe supports various types of DL architecture, such as CNN and RNN, initially targeting at image classification and segmentation.
\textbf{Torch} is a scientific computing framework based on the Lua programming language \cite{lua}, providing a rich collection of machine learning algorithms and data structures, such as tensors and corresponding mathematical operations, making it an ideal framework for DL. The goal of Torch is to achieve maximum flexibility and high speed. And notably, it comes with a large ecosystem with community supports. \cite{torch}
\textbf{Theano} is a Python library. It allows users to define, optimize, and evaluate DL models efficiently. It is tightly integrated with NumPy, a classical numerical library in Python and supports efficient symbolic differentiation. Theano also provides a rich high level APIs for DL. The design principle of Theano is to provide a computer algebra system (CAS) from the perspective of an optimizing compiler. \cite{theano} Thus, customized C codes of major mathematical operations can be generated by Theano to facilitate complex mathematical computation, e.g. DL, to approximate the performance of C programming language.~\cite{theano}

\section{Methodology and Baselines}
\label{methodology-baselines}
High-performance DL systems and applications demand both low latency and high accuracy simultaneously. With a large number of parameters in DL, it is very difficult to set these parameters, particularly the hyper-parameters, to achieve this goal.
We take a holistic approach to conduct a systematic empirical study on four representative DL frameworks: TensorFlow by Google~\cite{tensorflow}, Caffe by BVLC~\cite{caffe}, Torch~\cite{torch} by Lua community~\cite{lua}, and Theano (a Python library)~\cite{theano}.
As shown in Figure~\ref{fig:overview-DL-architecture}, the implementation and the runtime execution of different DL frameworks may differ in a number of ways, depending on their concrete implementation and configuration choices, including (1) the static components chosen as the parts of their runtime execution systems at the coding and implementation phase, such as the specific ML libraries, the specific parallel computing libraries, (2) the flexible components of their implementation systems, which are configurable prior to runtime, such as the concrete DNN structures, the hyper-parameters, such as the mini-batch size, the number of iterations, and the learning rate; and (3) the hardware platform compositions, such as with or without GPU devices, the number of GPU cards used, the type of network and storage runtime, such as using InfiniBand or Ethernet to connect computing nodes.  

Several existing research efforts have shown the impact of different hardware platforms on the performance of DL frameworks~\cite{shi-benchmarking-state-of-the-art, kim-performance-GPU, shams-evaluation-HPC, awan-caffe-variants}, and compared the performance of different DL frameworks with respect to their DNN structures and their default configuration settings~\cite{dawnbench, GT-ICDCS2018paper}. Thus, in this paper, we engage our empirical measurement study and comparison on characterization and analysis of DL frameworks in terms of how they respond to different configurations of their hyper-parameters, different types of datasets and different choices of parallel computing libraries. Both runtime performance in terms of training time and testing time and accuracy of DNN models produced by DL frameworks for prediction are measured in each set of experiments and reported with respect to their baseline configurations, which are the default settings of DNN structures and hyper-parameters recommended by the developers of each DL framework. We indicate those configuration settings of parallel computing libraries and hyper-parameters that can provide improved performance and/or improved accuracy over the recommended baseline default. 

The profiling tools used to measure the performance of the DL frameworks, in this study include the sysstat \cite{sysstat} and Nvidia SMI \cite{nvidia}. The sysstat is a collection of utilities to monitor system performance and usage, including the usage statistics for CPU and memory~\cite{sysstat}. The Nvidia SMI 
is a command line tool on top of the NVIDIA Management Library (NVML) for management and monitoring of NVIDIA GPU devices~\cite{nvidia}. 

\subsection{Hardware Platforms}
All experiments are primarily conducted on an Intel Xeon E5-1620 server (Server-1) with the 3.6Ghz 4-core CPU, DDR3 1600 MHz 8GB $\times$ 4 (32GB) memory, 256GB SSD, Nvidia GeForce GTX 1080 Ti with 3584 CUDA cores and 11 GB GDDR5X onboard memory, connected to the host via the PCIe 2.0 port, installed with Ubuntu 16.04 LTS, CUDA 8.0 and cuDNN 6.0. The same set of experiments were also repeated on a more powerful Intel Xeon server (Server-2), which has two Intel Xeon E5-2650 v4 2.2GHz CPUs, each with 12 physical cores, composing a NUMA architecture with DDR4 2400 MHz 16 GB$\times$12 (192GB) memory, 3TB SSD, 2 Nvidia Tesla M60, each has 2 GPUs, each GPU has 2048 CUDA cores and 8GB GDDR5 memory, installed with Ubuntu 16.04 LTS, CUDA 9.0 and cuDNN 7.0. Two M60 GPU cards are lined to the host via PCIe 3.0 ports. For both platforms, TensorFlow by default uses Eigen and the other DL frameworks are compiled with OpenBLAS. The default platform is the Intel Xeon E5-1620 server for all experiments reported in this paper unless stated otherwise. 

\subsection{Datasets} 

We choose three most popular and classic datasets: MNIST~\cite{mnistlenet}, CIFAR-10~\cite{cifar10-100}, and ImageNet (ILSVRC2012)\cite{imagenet-2015}. The first two datasets are the representative datasets used by almost all mainstream DL frameworks to tune and publish their default configurations. MNIST consists of 70,000 gray-scale images of ten handwritten digits, each image is $28\times28$ in size. CIFAR-10 consists of 60,000 colorful images of 10 classes, each is $32\times32$ in size. Table~\ref{table:datasets-size} shows the raw dataset size and its in-memory size for MNIST and CIFAR-10. The in-memory size is calculated with the assumption that the samples are stored in \textit{float32} matrices while labels are stored in \textit{int}. For ImageNet, ImageNet has 1,000 categories, the vast majority of its categories has about 1,300 samples for training, and each category owns 50 testing samples. The raw training dataset size is 140 GB while it is 6.3GB for its testing data.

\begin{table}[h!]
\vspace{-4mm}
\centering
\caption{\small{Datasets and their raw size \& in-memory}}
\label{table:datasets-size}
\vspace{-4mm}
\scalebox{0.85}{
\small
\begin{tabular}{|c|c|c|c|}
\hline
Dataset & Category & Raw Size (MB) & In-memory (MB) \\
\hline
\multirow{2}{*}{MNIST} & Training Samples & 44.92 & 179.67 \\
\cline{2-4} & Testing Samples & 7.48 & 29.95 \\
\hline
\multirow{2}{*}{CIFAR-10} & Training Samples & \multirow{2}{*}{162.17} & 195.50 \\
\cline{2-2} \cline{4-4} & Testing Samples & &  39.10 \\
\hline
\end{tabular}
} 
\end{table}

\begin{table}[h!]
\vspace{-4mm}
\centering
\caption{Default training parameters on MNIST}
\label{table:default-training-parameters-mnist}
\vspace{-4mm}
\scalebox{0.78}{
\small
\begin{tabular}{|c|c|c|c|c|}
    \hline
    Framework & TensorFlow & Caffe & Torch & Theano \\
    \hline
    Algorithm & Adam & SGD & SGD & SGD \\
    \hline
    \#Training Samples & 55,000 & 60,000 & 60,000 & 50,000 \\
    \hline
    \#Validation Samples & 5,000 & --- & --- & 10,000 \\
    \hline
    \#Testing Samples & 10,000 & 10,000 & 10,000 & 10,000 \\
    \hline
    Base Learning Rate & 0.0001 & 0.01 & 0.05 & 0.1 \\
    \hline
    Batch Size & 50 & 64 & 10 & 500 \\
    \hline
    \#Max Iterations & 20,000 & 10,000 & \textit{72,000} & \textit{20,000}\\
    \hline
    \#Epochs & \textit{18.18} & \textit{10.67} & 12 & 200 \\
    \hline
\end{tabular}
}
\vspace{-4mm}
\end{table}

\begin{table}[h!]
\centering
\caption{Default training parameters on CIFAR-10}
\label{table:default-training-parameters-cifar10}
\vspace{-4mm}
\scalebox{0.78}{
\small
\begin{tabular}{|c|c|c|c|c|}
    \hline
    Framework & TensorFlow & Caffe & Torch & Theano \\
    \hline
    Algorithm & SGD & SGD & SGD & SGD \\
    \hline
    \#Training Samples & 50,000 & 50,000 & 50,000 & 50,000 \\
    \hline
    \#Validation Samples & --- & --- & --- & --- \\
    \hline
    \#Testing Samples & 10,000 & 10,000 & 10,000 & 10,000 \\
    \hline
    Base Learning Rate & 0.1 & $0.001\rightarrow0.0001$ & 0.001 & 0.01 \\
    \hline
    Batch Size & 128 & 100 & 1 & 50 \\
    \hline
    \#Max Iterations & 1,000,000 & 5,000 & \textit{2,250,000} & \textit{50,000} \\
   \hline
    \#Epochs & \textit{2560} & \textit{8+2} & 45 & 50 \\
    \hline
\end{tabular}
}
\vspace{-6mm}
\end{table}

\begin{table*}[h!]
\vspace{-4mm}
\centering
\caption{Experimental Results on Server-1, using Default Settings by 4 frameworks}
\label{table:default-server-1}
\vspace{-7mm}
\subfloat[\footnotesize{Baseline Comparison on MNIST}]{
\scalebox{0.8}{
\small
\begin{tabular}{|c|c|c|c|}
\hline
Frameworks & Training Time (s) & Testing Time (s) & Accuracy (\%) \\ \hline
TF-CPU     & 1,114.34                                                    & {\bf 2.73}                                           & {\bf 99.24$\pm$0.05}                                                   \\ \hline
Caffe-CPU  & {\bf 512.18}                                   & 3.33                                                       & 99.04$\pm$0.02                                                   \\ \hline
Torch-CPU  & 9,647.34                                                    & 56.52                                                      & {\bf 99.24$\pm$0.00}                                                   \\ \hline
Theano-CPU & 11,555.43                                                   & 4.49                                                       & 99.08$\pm$0.00                                                   \\ \hline
TF-GPU     & {\bf 68.51}                & 0.26       & {\bf 99.21$\pm$0.03}                                                   \\ \hline
Caffe-GPU  & 97.02                      & 0.55            & 99.14$\pm$0.03                                                   \\ \hline
Torch-GPU  & 338.46                     & 1.73        & {\bf 99.22$\pm$0.00}                                                   \\ \hline
Theano-GPU & 560.04                  & {\bf 0.19}        & 99.05$\pm$0.00                                                   \\ \hline
\end{tabular}
\label{table:default-server-1-mnist}
} 
} 
\subfloat[\footnotesize{Baseline Comparison on CIFAR-10}]{
\scalebox{0.8}{
\small
\begin{tabular}{|c|c|c|c|}
\hline
Frameworks & Training Time (s) & Testing Time (s) & Accuracy (\%) \\ \hline
TF-CPU     & 219,169.14        & 4.80             & {\bf 86.90}         \\ \hline
Caffe-CPU  & 1,730.89          & 14.35            & 75.39         \\ \hline
Torch-CPU  & 54,830.26         & 114.48           & 66.20         \\ \hline
Theano-CPU & {\bf 646.9}        & {\bf 0.91}       & 56.04         \\ \hline
TF-GPU     & 12,477.05         & 2.34             & {\bf 87.00}         \\ \hline
Caffe-GPU  & 163.51            & 1.36             & 75.52         \\ \hline
Torch-GPU  & 1,906.56          & 3.77             & 65.96         \\ \hline
Theano-GPU & {\bf 105.27}            & {\bf 0.10}             & 54.49         \\ \hline
\end{tabular}
\label{table:default-server-1-cifar10}
} 
} 
\vspace{-3mm}
\end{table*}

\begin{table*}[h!]
\vspace{-4mm}
\centering
\caption{Experimental Results on Server-2, using Default Settings by 4 frameworks}
\label{table:default-server-2}
\vspace{-7mm}
\subfloat[\footnotesize{Baseline Comparison on MNIST}]{
\scalebox{0.8}{
\small
\begin{tabular}{|c|c|c|c|}
\hline
Frameworks & Training Time (s) & Testing Time (s) & Accuracy (\%) \\ \hline
TF-CPU     & 662.03    & 1.25 & {\bf 99.18$\pm$0.01} \\ \hline
Caffe-CPU  & {\bf 785.22}    & 2.72 & 99.11$\pm$0.04 \\ \hline
Torch-CPU  & 7,577.45  & 8.81 & 99.15$\pm$0.00 \\ \hline
Theano-CPU & 17,279.42 & 6.00 & 99.08$\pm$0.00 \\ \hline
TF-GPU     & 148.39    & 0.77 & \textbf{99.27}$\pm$0.04 \\ \hline
Caffe-GPU  & 135.74    & 0.67 & 99.13$\pm$0.01 \\ \hline
Torch-GPU  & 492.48    & 1.90 & {\bf 99.23$\pm$0.00} \\ \hline
Theano-GPU & 1,597.86  & 0.52 & 99.04$\pm$0.00 \\ \hline
\end{tabular}
\label{table:default-server-2-mnist}
} 
} 
\subfloat[\footnotesize{Baseline Comparison on CIFAR-10}]{
\scalebox{0.8}{
\small
\begin{tabular}{|c|c|c|c|}
\hline
Frameworks & Training Time (s) & Testing Time (s) & Accuracy (\%) \\ \hline
TF-CPU     & 119,102.62 & 2.69  & {\bf 87.00} \\ \hline
Caffe-CPU  & {\bf 2,708.88}   & {\bf 16.81} & 75.79 \\ \hline
Torch-CPU  & 46,850.68  & 52.10 & 66.20 \\ \hline
Theano-CPU & 532.81     & 1.17  & 49.66 \\ \hline
TF-GPU     & 26,160.37  & 5.76  & {\bf 86.00} \\ \hline
Caffe-GPU  & 183.49     & 1.09  & 75.86 \\ \hline
Torch-GPU  & 2,314.23   & 4.80  & 65.52 \\ \hline
Theano-GPU & 221.17     & 0.21  & 56.48 \\ \hline
\end{tabular}
\label{table:default-server-2-cifar10}
} 
} 
\vspace{-6mm}
\end{table*}

\subsection{Baseline Comparison}
Table~\ref{table:default-training-parameters-mnist} and Table~\ref{table:default-training-parameters-cifar10} show the primary hyper-parameter settings in the default configuration of TensorFlow, Caffe, Torch and Theano for MNIST and CIFAR-10 respectively, which are carefully chosen and recommended by the framework developers.
Given that the Theano release has only the default setting for MNIST \cite{theano}, we use a popular third-party Theano implementation on CIFAR-10~\cite{theano-cifar-10-tutorial}. To control the iterative training process, some DL frameworks use the notion of \#Epochs, such as Torch, Theano, whereas others use the notion of \#Iterations instead, such as TensorFlow, Caffe. We list both \#Epochs and \#Iterations for ease of comparison. For example, for MNIST, TensorFlow sets its max iterations to 20,000 and Caffe sets it to 10,000 while Torch and Theano set their \#Epochs to 12 or 200 respectively. TensorFlow and Theano reserve 5,000 and 10,000 samples respectively from the original training dataset as the \textit{validation samples}. Thus, the corresponding \#Epochs can be calculated as follows: 
\begin{equation}
\small
\#Epochs = batch\_size \times \#Iterations~/~\#Training~Samples
\label{formula:epoch-iteration}
\end{equation}
Thus, we obtain $\#Epochs = 50 \times 20,000 / (60,000 - 5,000) = 18.18$. Torch did not specify a default setting on \#Epochs for both datasets. We conduct experiments on Torch using Server-1 by varying \#Epochs on both datasets, and found that 12 epochs for MNIST and 45 epochs for CIFAR-10 are optimal \#Epochs and thus used as its baseline. 

From Table~\ref{table:default-training-parameters-mnist} and Table~\ref{table:default-training-parameters-cifar10}, the four DL frameworks adopt different hyper-parameters for
MNIST and CIFAR-10 datasets. For example, 
the SGD algorithm is used by all four frameworks for CIFAR-10 but TensorFlow sets Adam as its default optimizer for MNIST whereas Caffe, Torch and Theano still use SGD for MNIST. All four frameworks adopt different base learning rates (LR) and batch sizes for each of the two datasets. 
For TensorFlow, LR, batch size and \#Iterations increase significantly for CIFAR-10 compared to MNIST, with LR and \#Iterations 100$\times$ and 50$\times$ larger respectively. Similarly, Caffe increases its batch size to 100 and uses a two-phase training for CIFAR-10 to gain high accuracy, while Torch and Theano both decrease the LR and batch size in order to learn more elaborately.

In fact, the four DL frameworks also set their default DNN structures differently for MNIST and CIFAR-10.
The four frameworks adopt a similar DNN structure with 2 convolution layers for MNIST but use significantly different settings for CIFAR-10. 
These baseline comparisons imply that the performance and accuracy of DL frameworks can be sensitive to the type of datasets and the specific configurations of their hyper-parameters. The sensitivity to datasets leads to different optimal settings of the hyper-parameters for different datasets under the same DL framework (data dependency). The sensitivity to the specific implementation choices (recall 
Figure~\ref{fig:overview-DL-architecture}) leads to the divergence of different DL frameworks on the optimal settings of the hyper-parameters for the same dataset (framework dependency). 

\subsection{Empirical Study Objectives} 
The above baseline comparison and analysis motivate us to conduct empirical measurement and comparative study on the four DL frameworks with three complementary objectives. First, we will compare the effectiveness and efficiency of the four DL frameworks using their default baseline configurations for MNIST and CIFAR-10 on different configurations of hardware and parallel computing libraries. Second, we will characterize and analyze the optimal settings of those hyper-parameters that are critical to both performance and accuracy, including both the impact of individual hyper-parameters and the impact of the combo-settings of two or more hyper-parameters. Third, we will measure and characterize CPU and memory resource consumptions by different DL frameworks for different hardware configurations, parallel computing libraries and hyper-parameter settings. We believe that this comprehensive empirical characterization and analysis of the DL frameworks can serve as a practical guidance for researchers, developers, domain scientists and big data professionals in both choosing a DL framework most effective for their big datasets and learning tasks, and tuning the DL frameworks for high performance and high accuracy.

\begin{table*}[!h]
\vspace{-2mm}
\centering
\caption{Varying \#GPUs (Server-2)}
\label{table:varying-gpus}
\vspace{-4mm}
\small
\scalebox{0.8}{
\begin{tabular}{|c|c|c|c|c|c|c|c|c|}
\hline
\multirow{2}{*}{Framework}  & \multirow{2}{*}{Dataset}  & \multirow{2}{*}{\#GPUs} & \multirow{2}{*}{\#Iterations} & \multicolumn{2}{c|}{Batch Size} & \multirow{2}{*}{\begin{tabular}[c]{@{}c@{}}Training\\ Time (s)\end{tabular}} & \multirow{2}{*}{\begin{tabular}[c]{@{}c@{}}Testing\\ Time (s)\end{tabular}} & \multirow{2}{*}{\begin{tabular}[c]{@{}c@{}}Accuracy\\ (\%)\end{tabular}} \\ \cline{5-6}
                            &                           &                         &                               & Total         & Per GPU         &                                                                              &                                                                             &                                                                          \\ \hline
\multirow{3}{*}{TF} & \multirow{3}{*}{ImageNet} & 1                       & 100000                        & 32            & 32              & \DEFAULTHIGHLIGHT{111,781.20}                                                                   & \DEFAULTHIGHLIGHT{429.11}                                                                      & \DEFAULTHIGHLIGHT{45.09}                                                                    \\ \cline{3-9} 
                            &                           & 2                       & 100000                        & 64            & 32              & 118,972.30                                                                   & 427.68                                                                      & 56.66                                                                    \\ \cline{3-9} 
                            &                           & 4                       & 100000                        & 128           & 32              & 131,796.70                                                                   & \textbf{426.46}                                                                      & \textbf{63.81}                                                                    \\ \hline
\multirow{6}{*}{Caffe}      & \multirow{3}{*}{MNIST}    & 1                       & 10000                         & 64            & 64              & \DEFAULTHIGHLIGHT{135.74}    & \DEFAULTHIGHLIGHT{0.67} & \DEFAULTHIGHLIGHT{99.13$\pm$0.01}                                                                    \\ \cline{3-9} 
                            &                           & 2                       & 10000                         & 128           & 64              & 146.67                                                                       & 0.68                                                                        & 99.02                                                                    \\ \cline{3-9} 
                            &                           & 4                       & 10000                         & 256           & 64              & 155.80                                                                       & 0.68                                                                        & 99.10                                                                    \\ \cline{2-9} 
                            & \multirow{3}{*}{CIFAR-10} & 1                       & 5000                          & 100           & 100             & \DEFAULTHIGHLIGHT{184.19}                                                                       & \DEFAULTHIGHLIGHT{1.10}                                                                        & \DEFAULTHIGHLIGHT{75.47}                                                                    \\ \cline{3-9} 
                            &                           & 2                       & 5000                          & 200           & 100             & 188.74                                                                       & 1.10                                                                        & 76.62                                                                    \\ \cline{3-9} 
                            &                           & 4                       & 5000                          & 400           & 100             & 192.73                                                                       & 1.10                                                                        & \textbf{77.27}                                                                    \\ \hline
\end{tabular}
} 
\vspace{-6mm}
\end{table*}

\section{Experimental Comparison and Analysis}
\label{section:experiments}

\subsection{Impact of Hardware Configurations} 

\subsubsection{Baseline Experiments} \label{section:baseline-experiments}

We first compare the baseline performance of all four DL frameworks using their default settings on MNIST and CIFAR-10 on both CPU and GPU, and the results are shown in Table \ref{table:default-server-1} on Server-1 and Table \ref{table:default-server-2} on Server-2. We also show the confidence interval (mean$\pm$stddev) derived from 5 repeated experiments for the accuracy on MNIST due to their similar accuracy. We make four interesting observations.

{\em First}, 
regarding accuracy, under both CPU and GPU on Server-1 and Server-2, TensorFlow (TF) and Torch achieved higher accuracy for MNIST, even though Torch has smaller \#Epochs, kernel size and batch size than TensorFlow. But for CIFAR-10, TensorFlow has the highest accuracy, followed by Caffe, then Torch with Theano the worst, though Theano spent the least training and testing time. Also, for CIFAR-10, Torch-GPU has a slightly lower accuracy (65.96\%) than Torch-CPU (66.20\%), one reason might be that Torch uses SpatialConvolutionMap on CPU and uses SpatialConvolutionMM on GPU due to the lack of the corresponding implementation on GPU \cite{torch}. Overall, all frameworks achieve much lower accuracy on the content-rich CIFAR-10 dataset compared to the simple gray-scale MNIST dataset, due to the low entropy of MNIST, thus easier DNN feature extraction. Specifically, for MNIST, all frameworks obtain an accuracy of $>99\%$. While, for CIFAR-10, all frameworks achieve an accuracy of $\leq 87.00\%$.

{\em Second}, 
regarding training and testing time, the GPU significantly shortens the training and testing time for all frameworks on both datasets and two servers, except that Theano is slightly lower, but offers higher accuracy in Caffe for both datasets than CPU. For CIFAR-10, the highest accuracy of TensorFlow is at the cost of taking significantly longer training time on both CPU and GPU.

{\em Third}, for CIFAR-10, Torch-CPU and Theano-CPU achieved higher accuracy of 66.20\% and 56.04\% respectively than Torch-GPU (65.96\%) and Theano-GPU (54.49\%) on Server-1. This observation indicates that even GPU accelerates the training and testing, it may not lead to high accuracy. One primariy reason is that DL frameworks are implemented with different parallel computing libraries on CPU (e.g., OpenBLAS) and GPU (e.g., CUDA) is because the accuracy and performance may depend on multiple factors, including the specific compositions of hardware and different layers of software, ranging from numerical library to the parallel computation library.

{\em Fourth}, comparing the memory and CPU of the two servers, Server-2 has much larger memory (192GB) and higher performance CPU (E5-2650 v4) than Server-1 (32GB, E5-1620). However, Caffe achieved higher runtime performance on Server-1 than Server-2 for both MNIST and CIFAR-10, i.e., 512.18s (Server-1) and 785.22s (Server-2) for training on MNIST. Theano also has obtained shorter training and testing time and achieved higher accuracy on Server-1 than Server-2 for MNIST. These observations indicate that higher capacity of memory and CPU may not result in shorter training/testing time and better training accuracy and we conjecture that scalable advancement in DL software frameworks that can leverage more powerful hardware capabilities is an important and critical challenge for scaling DLaaS for big data powered DL workloads.

\subsubsection{Impact of Varying \#GPUs}
DL frameworks continue to improve over time for better support of multiple GPUs. TensorFlow comes with a default implementation for ImageNet with multi-GPU support. Caffe inherently supports multiple GPUs (enabled for this set of experiments). However, TensorFlow, Torch and Theano do not implement their support for multiple GPUs for MNIST and CIFAR-10 in their default configurations. 
Thus, we conduct a set of experiments to show the impact of \#GPUs by varying it from 1, 2 to 4 on the runtime performance and accuracy of TensorFlow and Caffe. For TensorFlow, the neural network structure is the Inception-v3~\cite{inception-v3}, and it costs 5 billion multiply-adds per inference and uses almost 25 million parameters. We set the batch size for each GPU to be 32 as recommended in~\cite{inception-v3} and other parameters set as default. For Caffe, the parameters are set as default with a fixed batch size per GPU, 64 for MNIST and 100 for CIFAR-10.

Table \ref{table:varying-gpus} shows the experimental results. We use the red font to show the default setting while the black bold face shows the empirical optimal configuration
found in our study. We observe that the training time increases as the batch size grows for both TensorFlow and Caffe. This is primarily due to the synchronization overhead of multiple GPUs. The testing time remains similar because the testing is performed only on one GPU. Furthermore, the accuracy of TensorFlow on ImageNet and the accuracy of Caffe on CIFAR-10 both increase as the total batch size increases given the fixed \#Iterations. However, the accuracy of Caffe on MNIST fluctuates when we vary the number of GPUs, and also the accuracy of Caffe on MNIST with 2 or 4 GPUs are lower than its accuracy with 1 GPU. This shows that more GPUs for MNIST may not result in shorter training time and higher accuracy. The possible explanation is that multiple GPUs may have a higher adverse impact on the numeric precision of the training on MNIST than the other two datasets due to different dataset characteristics and DNN models, therefore, it results in the accuracy fluctuation. This set of experiments also shows that for content-rich and color-rich datasets like CIFAR-10 and ImageNet, more GPUs can improve the accuracy of trained DNN models, indicating that larger batch sizes bundled with multi-GPUs hold the potential to improve the accuracy with little cost of training time, particularly for content-rich datasets. Thus, developing dataset feature aware configuration of GPUs and deep neural network structures can be a complimentary dimension for benchmarking and performance tuning.


\begin{table}[h!]
\vspace{-3mm}
\centering
\caption{\small{CPU $>$ GPU (MNIST)}}
\label{table:cpu-vs-gpu}
\vspace{-5mm}
\small
\scalebox{0.71}{
\begin{tabular}{|c|c|c|c|c|}
\hline
Framework               & Setting             & Training Time (s) & Testing Time (s) & Accuracy (\%) \\ \hline
\multirow{3}{*}{Caffe}  & GPU-1               & \DEFAULTHIGHLIGHT{97.02}             & \DEFAULTHIGHLIGHT{0.55}             & \DEFAULTHIGHLIGHT{99.14$\pm$0.03}         \\ \cline{2-5} 
                        & GPU-2               & \DEFAULTHIGHLIGHT{135.74}            & \DEFAULTHIGHLIGHT{0.67}             & \DEFAULTHIGHLIGHT{99.13$\pm$0.01}         \\ \cline{2-5} 
                        & MKL (CPU-2) & \textbf{59.11}             & \textbf{0.25}             & \textbf{99.18}         \\ \hline
\multirow{2}{*}{Theano} & GPU-2               & \DEFAULTHIGHLIGHT{1,597.86}          & \DEFAULTHIGHLIGHT{0.52}             & \DEFAULTHIGHLIGHT{99.04$\pm$0.00}         \\ \cline{2-5} 
                        & MKL (CPU-2) & \textbf{691.75}            & \textbf{0.28}             & 98.99         \\ \hline
\end{tabular}
} 
\vspace{-4mm}
\end{table}

\subsubsection{CPU v.s. CPU with GPU}
Our baseline experiments have shown that the hardware configuration of CPU with GPUs typically outperforms the CPU server without GPU by an order of magnitude, denoted by CPU $<$ GPU for presentation brevity. However, our empirical measurements have shown that in some scenarios, CPU without GPUs, with proper optimizations, may outperform the CPU with GPUs in terms of training and testing time as well as accuracy, denoted by CPU $>$ GPU. Table \ref{table:cpu-vs-gpu} shows the results of one such set of measurements. GPU-1 and GPU-2 denote the default GPU configurations on Server-1 and Server-2 while on Server-2, we replace OpenBLAS, the default for Caffe and Theano on CPU with MKL as the parallel computing library, denoted as MKL (CPU-2). We observe that the performance of Caffe with MKL (CPU-2) shows significant improvement over GPU-1 and GPU-2. Also the training time Caffe with MLK on Server-2 improves over that with GPU-1 by 1.64$\times$ and GPU-2 by 2.30$\times$. Furthermore, Caffe achieved higher accuracy of 99.18\%, slightly higher than GPU-1 (99.14\%) and GPU-2 (99.13\%). Similar observations are made for Theano. With MKL, Theano MKL (CPU-2) also achieved higher runtime performance than on GPU-2 with the training time and testing time reduced from 1,597.86s and 0.52s (GPU-2) to 691.75s and 0.28s respectively. These observations show the opportunities and the potentials of parallel computation optimizations for CPU without GPUs. It also shows the potential of better DL software frameworks for a given hardware configuration platform, be it CPU or CPU with GPUs.


\begin{table*}[h!]
\vspace{-4mm}
\centering
\caption{Default Parallel Computing Libraries v.s. MKL (Server-1, MNIST)}
\label{table:default-math-lib-mkl-1}
\vspace{-4mm}
\scalebox{0.8}{
\makebox[\linewidth][c]{
\small
\begin{tabular}{|c|c|c|c|c|c|c|c|c|c|c|}
\hline
\multirow{2}{*}{Framework}      & \multirow{2}{*}{Setting} & \multirow{2}{*}{\begin{tabular}[c]{@{}c@{}}Training\\ Time (s)\end{tabular}} & \multirow{2}{*}{\begin{tabular}[c]{@{}c@{}}Testing\\ Time (s)\end{tabular}} & \multirow{2}{*}{\begin{tabular}[c]{@{}c@{}}Accuracy\\ (\%)\end{tabular}} & \multicolumn{4}{c|}{CPU Usage (\% AVG)} & \multicolumn{2}{c|}{Memory (MB)} \\ \cline{6-11} 
                                &                          &                                                                              &                                                                             &                                                                          & user        & system      & iowait & total      & AVG             & MAX            \\ \hline
\multirow{2}{*}{TensorFlow-CPU} & Eigen                    & \DEFAULTHIGHLIGHT{1,114.34}                                                                     & \DEFAULTHIGHLIGHT{2.73}                                                                        & \DEFAULTHIGHLIGHT{99.24$\pm$0.05}                                                                    & \DEFAULTHIGHLIGHT{72.26}       & \DEFAULTHIGHLIGHT{7.88}        & \DEFAULTHIGHLIGHT{0.01}  &  \DEFAULTHIGHLIGHT{80.15}     & \DEFAULTHIGHLIGHT{736.73}          & \DEFAULTHIGHLIGHT{6268.55}        \\ \cline{2-11} 
                                & MKL              & 6,100.86                                                                     & 3.51                                                                        & 99.21                                                                    & 47.47       & 51.65       & 0.00   & \textbf{99.12}     & \textbf{672.81}          & \textbf{2762.46}        \\ \hline
\multirow{2}{*}{Caffe-CPU}      & OpenBLAS          & \DEFAULTHIGHLIGHT{512.18}                                                                       & \DEFAULTHIGHLIGHT{3.33}                                                                        & \DEFAULTHIGHLIGHT{99.04$\pm$0.02}                                                                    & \DEFAULTHIGHLIGHT{49.59}       & \DEFAULTHIGHLIGHT{47.52}       & \DEFAULTHIGHLIGHT{0.01}   & \DEFAULTHIGHLIGHT{97.12}     & \DEFAULTHIGHLIGHT{464.91}          & \DEFAULTHIGHLIGHT{491.05}         \\ \cline{2-11} 
                                & MKL              & \textbf{187.33}                                                                       & \textbf{0.75}                                                                        & \textbf{99.13}                                                                    & 47.90       & \textbf{0.27}        & 0.01  & 48.18      & \textbf{163.11}          & \textbf{180.48}         \\ \hline
\multirow{2}{*}{Torch-CPU}      & OpenBLAS          & \DEFAULTHIGHLIGHT{9,647.34}                                                                     & \DEFAULTHIGHLIGHT{56.52}                                                                       & \DEFAULTHIGHLIGHT{99.24$\pm$0.00}                                                                    & \DEFAULTHIGHLIGHT{34.64}       & \DEFAULTHIGHLIGHT{65.20}       & \DEFAULTHIGHLIGHT{0.00}  & \DEFAULTHIGHLIGHT{99.84}      & \DEFAULTHIGHLIGHT{661.51}          & \DEFAULTHIGHLIGHT{768.32}         \\ \cline{2-11} 
                                & MKL              & \textbf{1,482.33}                                                                     & \textbf{3.05}                                                                        & 99.17                                                                    & 31.39       & \textbf{13.76}       & 0.00  & 45.15      & \textbf{439.13}          & \textbf{495.88}         \\ \hline
\multirow{2}{*}{Theano-CPU}     & OpenBLAS          & \DEFAULTHIGHLIGHT{11,555.43}                                                                    & \DEFAULTHIGHLIGHT{4.49}                                                                        & \DEFAULTHIGHLIGHT{99.08$\pm$0.00}                                                                    & \DEFAULTHIGHLIGHT{41.16}       & \DEFAULTHIGHLIGHT{44.33}       & \DEFAULTHIGHLIGHT{0.00}  & \DEFAULTHIGHLIGHT{85.49}      & \DEFAULTHIGHLIGHT{1621.72}         & \DEFAULTHIGHLIGHT{4062.09}        \\ \cline{2-11} 
                                & MKL              & \textbf{7,137.35}                                                                     & \textbf{2.75}                                                                        & 99.08                                                                    & 31.60       & \textbf{0.70}        & 0.01  & 32.31      & \textbf{814.32}          & \textbf{937.92}         \\ \hline
\end{tabular}
}}
\vspace{-4mm}
\end{table*}

\begin{table*}[h!]
\centering
\caption{Default Parallel Computing Libraries v.s. MKL (Server-2, MNIST)}
\label{table:default-math-lib-mkl-2}
\vspace{-4mm}
\scalebox{0.8}{
\makebox[\linewidth][c]{
\small
\begin{tabular}{|c|c|c|c|c|c|c|c|c|c|c|}
\hline
\multirow{2}{*}{Framework}      & \multirow{2}{*}{Setting} & \multirow{2}{*}{\begin{tabular}[c]{@{}c@{}}Training\\ Time (s)\end{tabular}} & \multirow{2}{*}{\begin{tabular}[c]{@{}c@{}}Testing\\ Time (s)\end{tabular}} & \multirow{2}{*}{\begin{tabular}[c]{@{}c@{}}Accuracy\\ (\%)\end{tabular}} & \multicolumn{4}{c|}{CPU Usage (\% AVG)} & \multicolumn{2}{c|}{Memory (MB)} \\ \cline{6-11} 
                                &                          &                                                                              &                                                                             &                                                                          & user        & system      & iowait  & total   & AVG             & MAX            \\ \hline
\multirow{2}{*}{TensorFlow-CPU} & Eigen           & \DEFAULTHIGHLIGHT{662.03}    & \DEFAULTHIGHLIGHT{1.25} & \DEFAULTHIGHLIGHT{99.18$\pm$0.01} & \DEFAULTHIGHLIGHT{34.20} & \DEFAULTHIGHLIGHT{15.83} & \DEFAULTHIGHLIGHT{0.00} & \DEFAULTHIGHLIGHT{50.03} & \DEFAULTHIGHLIGHT{1079.77} & \DEFAULTHIGHLIGHT{2703.15} \\ \cline{2-10} 
                                & MKL     & 3,364.51  & 1.12 & 99.14 & \textbf{34.48} & 64.99 & 0.00 & \textbf{99.47} & 1324.39 & 3342.94 \\ \hline
\multirow{2}{*}{Caffe-CPU}      & OpenBLAS & \DEFAULTHIGHLIGHT{785.22}    & \DEFAULTHIGHLIGHT{2.72} & \DEFAULTHIGHLIGHT{99.11$\pm$0.04} & \DEFAULTHIGHLIGHT{24.75} & \DEFAULTHIGHLIGHT{66.61} & \DEFAULTHIGHLIGHT{0.00} & \DEFAULTHIGHLIGHT{91.36} & \DEFAULTHIGHLIGHT{318.71}  & \DEFAULTHIGHLIGHT{341.53}  \\ \cline{2-11} 
                                & MKL     & \textbf{59.11}     & \textbf{0.25} & \textbf{99.18} & \textbf{43.54} & 0.89  & 0.00 & 44.43 & \textbf{62.55}   & \textbf{80.18}   \\ \hline
\multirow{2}{*}{Torch-CPU}      & OpenBLAS & \DEFAULTHIGHLIGHT{7,577.45}  & \DEFAULTHIGHLIGHT{8.81} & \DEFAULTHIGHLIGHT{99.15$\pm$0.00} & \DEFAULTHIGHLIGHT{27.35} & \DEFAULTHIGHLIGHT{71.19} & \DEFAULTHIGHLIGHT{0.00} & \DEFAULTHIGHLIGHT{98.54} & \DEFAULTHIGHLIGHT{569.68}  & \DEFAULTHIGHLIGHT{719.46}  \\ \cline{2-11} 
                                & MKL     & \textbf{1,555.17}  & \textbf{5.53} & \textbf{99.22} & 9.95  & \textbf{2.04}  & 0.00 & 11.99 & \textbf{429.67}  & \textbf{480.41}  \\ \hline
\multirow{2}{*}{Theano-CPU}     & OpenBLAS & \DEFAULTHIGHLIGHT{17,279.42} & \DEFAULTHIGHLIGHT{6.00} & \DEFAULTHIGHLIGHT{99.08$\pm$0.00} & \DEFAULTHIGHLIGHT{25.20} & \DEFAULTHIGHLIGHT{69.29} & \DEFAULTHIGHLIGHT{0.00} & \DEFAULTHIGHLIGHT{94.49} & \DEFAULTHIGHLIGHT{843.65}  & \DEFAULTHIGHLIGHT{1139.44} \\ \cline{2-11} 
                                & MKL     & \textbf{691.75}    & \textbf{0.28} & 98.99 & \textbf{41.75} & \textbf{0.80}  & 0.00 & 42.55 & \textbf{550.44}  & \textbf{574.49}  \\ \hline
\end{tabular}
}}
\vspace{-4mm}
\end{table*}

\subsection{Impact of Parallel Computing Libraries} \label{section:impact-parallel-computing-libraries}

The bagging schemes enable the partitioning of large datasets into many small mini-batches such that massive parallel processing is possible for deep neural networks (DNNs) with huge feature maps. As a result, all DL frameworks are compute-intensive. The runtime performance of deep learning frameworks highly relies on the performance of the underlying parallel computing libraries, as shown in the previous set of experiments. In this section, we primarily study the impact of different parallel computing libraries and the configurations on both the runtime performance and the accuracy of different DL frameworks. For the two hardware platforms (Server-1 and Server-2), the DL frameworks are compiled natively with Eigen or OpenBLAS or MKL. Eigen is only used as the default by TensorFlow while Caffe, Torch and Theano use OpenBLAS as their default parallel computation library. Eigen is tightly integrated into TensorFlow~\cite{tensorflow}, while OpenBLAS is a popular open-source, optimized BLAS (Basic Linear Algebra Subprograms) library. MKL is developed by Intel to better utilize the computation power of Intel CPUs for optimal computing performance. This set of experiments compares the use of MKL as the parallel computation library with the use of the default parallel computing libraries in all four DL frameworks on both Server-1 and Server-2. We also vary the environmental variable OPENBLAS\_NUM\_THREADS in OpenBLAS (abbreviated as \#THREADS) to set the number of threads involved in parallel computation (default: unset) on Caffe, Torch and Theano for Serve-2. Two key system performance indicators, CPU usage and memory utilization, are also measured in this set of experiments.

\subsubsection{Impact of Parallel Computing Libraries}

Table \ref{table:default-math-lib-mkl-1} and Table \ref{table:default-math-lib-mkl-2} show the measurement results by using different parallel computing libraries on Server-1 and Server-2 respectively. We make two interesting observations. 

{\em First}, since MKL is well-optimized by Intel for Intel CPUs used by both Server-1 and Server-2. We observe indeed the better performance for MKL compared to the scenarios where OpenBLAS is used as the default parallel computing library, such as Caffe, Torch, and Theano. Although MKL achieved much higher runtime performance than OpenBLAS in Caffe, Torch and Theano, TensorFlow with Eigen as its default parallel computation library significantly outperforms MKL for training on Server-1 and Server-2 by 5.47$\times$ and 5.08$\times$ respectively. We conjecture that the high performance of Eigen in TensorFlow is due to their tight integration, even though Eigen may not be properly optimized for Intel CPUs as MKL does, replacing Eigen with MKL in TensorFlow leads to significant performance reduction for TensorFlow. However, the choice of Eigen or MKL has low impact on the accuracy of TensorFlow trained DNN models. Also the accuracy for an individual DL framework with different parallel computing libraries varies within 0.10\% on both Server-1 and Server-2. This observation manifests the feasibility of changing the parallel computing libraries to attain better runtime performance with negligible impact on accuracy.


{\em Second}, We measure the total CPU usage as the overall utilization of CPU. High CPU utilization typically implies better runtime performance. However, the highest total CPU usage of DL frameworks does not correspond to the highest runtime performance because the performance of the parallel computation library for an DL framework is a dominating performance factor. For TensorFlow, its total CPU usage with MKL is 99.12\% on Server-1 and 99.47\% on Server-2, which is much higher than Eigen with 80.15\% on Server-1 and 50.03\% on Server-2. However, TensorFlow with MKL has much worse training performance compared to TensorFlow with Eigen. Moreover, Caffe, Torch and Theano with OpenBLAS (their default option) achieved the highest total CPU usage on both Server-1 and Server-2 respectively, but suffer from the worst training time. Meanwhile, it is also observed that the shorter training often corresponds to lower memory usage, possibly corresponding to smaller memory footprints.

\begin{table*}[h!]
\vspace{-3mm}
\centering
\caption{Varying \#THREADS with OpenBLAS (Server-2, MNIST)}
\label{table:openblas-threads}
\vspace{-4mm}
\scalebox{0.8}{
\begin{tabular}{|c|c|c|c|c|c|c|c|c|c|c|}
\hline
\multirow{2}{*}{Framework} & \multirow{2}{*}{\#THREADS} & \multirow{2}{*}{\begin{tabular}[c]{@{}c@{}}Training\\ Time (s)\end{tabular}} & \multirow{2}{*}{\begin{tabular}[c]{@{}c@{}}Testing\\ Time (s)\end{tabular}} & \multirow{2}{*}{\begin{tabular}[c]{@{}c@{}}Accuracy\\ (\%)\end{tabular}} & \multicolumn{4}{c|}{CPU Usage (\% AVG)} & \multicolumn{2}{c|}{Memory (MB)} \\ \cline{6-11} 
                           &                            &                                                                              &                                                                             &                                                                          & user    & system   & iowait   & total   & AVG            & MAX             \\ \hline
\multirow{7}{*}{Caffe-CPU}     & 4                          & 617.59                                                                       & 3.47                                                                        & \textbf{99.14}                                                                    & 5.64    & \textbf{2.66}     & 0        & 8.30     & 380.24         & 405.90           \\ \cline{2-11} 
                           & 8                          & 576.93                                                                       & 3.32                                                                        & 99.06                                                                    & 9.36    & 7.15     & 0        & 16.51   & \textbf{309.41}         & \textbf{325.82}          \\ \cline{2-11} 
                           & 12                         & 640.40                                                                        & 3.78                                                                        & 99.10                                                                     & 13.81   & 10.53    & 0        & 24.34   & 311.34         & 653.50           \\ \cline{2-11} 
                           & 16                         & 594.15                                                                       & 2.88                                                                        & 99.11                                                                    & 14.28   & 18.42    & 0        & 32.70    & 307.33         & 592.98          \\ \cline{2-11} 
                           & 24                         & \textbf{543.92}                                                                       & 2.85                                                                        & 99.12                                                                    & 14.90    & 34.10     & 0        & 49.00      & 320.42         & 619.70           \\ \cline{2-11} 
                           & 48                         & 785.32                                                                       & 2.85                                                                        & 99.09                                                                    & \textbf{24.83}   & 67.14    & 0        & \textbf{91.97}   & 316.74         & 336.86          \\ \cline{2-11} 
                           & unset                      & \DEFAULTHIGHLIGHT{785.22}                                                                       & \DEFAULTHIGHLIGHT{2.72}                                                                        & \DEFAULTHIGHLIGHT{99.11$\pm$0.04}                                                                       & \DEFAULTHIGHLIGHT{24.75}   & \DEFAULTHIGHLIGHT{66.61}    & \DEFAULTHIGHLIGHT{0}        & \DEFAULTHIGHLIGHT{91.36}   & \DEFAULTHIGHLIGHT{318.71}         & \DEFAULTHIGHLIGHT{341.53}          \\ \hline
\multirow{7}{*}{Torch-CPU}     & 4                          & \textbf{1,224.94}                                                                     & 8.89                                                                        & 99.13                                                                    & 9.14    & \textbf{3.36}     & 0        & 12.5    & 335.97         & 415.75          \\ \cline{2-11} 
                           & 8                          & 1,281.14                                                                     & 9.27                                                                        & 99.17                                                                    & 12.25   & 8.57     & 0        & 20.82   & 368.76         & 439.82          \\ \cline{2-11} 
                           & 12                         & 1,231.05                                                                     & 8.72                                                                        & \textbf{99.21}                                                                    & 12.18   & 15.00       & 0        & 27.18   & 368.15         & \textbf{404.39}          \\ \cline{2-11} 
                           & 16                         & 1,246.37                                                                     & \textbf{8.56}                                                                        & 99.20                                                                     & 14.33   & 20.57    & 0        & 34.90    & 345.14         & 409.06          \\ \cline{2-11} 
                           & 24                         & 1,339.94                                                                     & 8.68                                                                        & 99.17                                                                    & 18.08   & 32.50     & 0        & 50.58   & \textbf{328.93}         & 416.78          \\ \cline{2-11} 
                           & 48                         & 7,591.54                                                                     & 8.91                                                                        & 99.15                                                                    & \textbf{27.37}   & 71.10     & 0        & 98.47   & 447.64         & 608.09          \\ \cline{2-11} 
                           & unset                      & \DEFAULTHIGHLIGHT{7,577.45}                                                                     & \DEFAULTHIGHLIGHT{8.81}                                                                        & \DEFAULTHIGHLIGHT{99.15$\pm$0.00}                                                                    & \DEFAULTHIGHLIGHT{27.35}   & \DEFAULTHIGHLIGHT{71.19}    & \DEFAULTHIGHLIGHT{0}        & \DEFAULTHIGHLIGHT{98.54}   & \DEFAULTHIGHLIGHT{569.68}         & \DEFAULTHIGHLIGHT{719.46}          \\ \hline
\multirow{7}{*}{Theano-CPU}    & 4                          & 15,905.31                                                                    & 5.81                                                                        & 99.08                                                                    & 4.85    & \textbf{3.44}     & 0        & 8.29    & 853.52         & 1,197.63        \\ \cline{2-11} 
                           & 8                          & \textbf{15,117.45}                                                                    & 5.54                                                                        & 99.08                                                                    & 7.18    & 8.11     & 0        & 15.29   & 836.19         & \textbf{1,168.83}        \\ \cline{2-11} 
                           & 12                         & 16,183.43                                                                    & 5.62                                                                        & 99.08                                                                    & \textbf{10.10}    & 12.74    & 0        & 22.84   & 857.61         & 1,200.20        \\ \cline{2-11} 
                           & 16                         & 16,785.99                                                                    & 5.47                                                                        & 99.08                                                                    & 7.84    & 11.75    & 0        & 19.59   & 851.39         & 1,208.58        \\ \cline{2-11} 
                           & 24                         & 15,289.40                                                                    & \textbf{5.41}                                                                        & 99.08                                                                    & 6.30     & 13.89    & 0        & 20.19   & \textbf{823.22}         & 1,154.83        \\ \cline{2-11} 
                           & 48                         & 17,312.62                                                                    & 5.98                                                                        & 99.08                                                                    & 9.63    & 28.08    & 0        & 37.71   & 860.70          & 1,234.82        \\ \cline{2-11} 
                           & unset                      & \DEFAULTHIGHLIGHT{17,264.30}                                                                    & \DEFAULTHIGHLIGHT{6.00}                                                                           & \DEFAULTHIGHLIGHT{99.08$\pm$0.00}                                                                    & \DEFAULTHIGHLIGHT{9.67}    & \DEFAULTHIGHLIGHT{28.18}    & \DEFAULTHIGHLIGHT{0}        & \DEFAULTHIGHLIGHT{37.85}   & \DEFAULTHIGHLIGHT{864.37}         & \DEFAULTHIGHLIGHT{1,232.59}        \\ \hline
\end{tabular}
}
\vspace{-4mm}
\end{table*}

\subsubsection{Impact of \#THREADS in Parallel Computing}

Server-2 has 24 physical cores, that is 48 logical cores with HT (Hyper-threading) enabled. Table \ref{table:openblas-threads} shows the experimental results on Sever-2 by varying the \#THREADS from 4 to 48 of OpenBLAS for Caffe, Torch and Theano. The experimental results further confirm our previous observations on trivial impact of specific parallel computing libraries on the accuracy. We observe small accuracy difference by varying \#THREADS for each of these three DL frameworks, i.e., the accuracy difference is within 0.08\% for Caffe, 0.06\% for Torch and no difference for Theano, demonstrating the tiny impact of different thread count configurations in the OpenBLAS parallel computing library. For CPU usage, as the \#THREADS increases, the total CPU usage and the percentage of CPU usage for systems also increase. Hence, the performance degradation with larger \#THREADS is likely caused by the increased overhead of thread synchronization. In addition, the optimal setting of the \#THREADS is 24 for Caffe, 4 for Torch, and 8 for Theano on the same hardware platform, showing that the implementations of DL software frameworks may have larger impact on the parallel computing performance. In particular, Torch with \#TREAHDS=4 in OpenBLAS achieved shorter training time (1,224.94s) than using Intel optimized MKL (1,555.17s), demonstrating that OpenBLAS with a proper configuration could outperform MKL. In summary, tight integration of parallel computing library and its configurations with DL frameworks, such as TensorFlow with Eigen, Torch with OpenBLAS optimal \#TREAHDS configuration, are highly recommended for performance tuning of DL frameworks.

\subsection{Impact of Hyper-parameter: \#Epochs (\#Iterations)}
\label{subsection:impact-epochs-iterations}

Recall the comparative analysis of the baseline configurations of DL frameworks presented earlier, we compared their training time performance and accuracy based on their default settings for hyper-parameters and DNN structures. In this section, we run a set of experiments and tune the settings of individual hyper-parameters to identify those optimal settings for each of the four DL frameworks. We first study the impact of \#Epochs (or \#Iterations) and keep the framework dependent and dataset dependent default settings for other parameters, as those shown in Table \ref{table:default-training-parameters-mnist} and Table \ref{table:default-training-parameters-cifar10}.

\subsubsection{TensorFlow}
From the baseline experiments on CIFAR-10 on both CPU and GPU, TensorFlow shows the highest accuracy at the cost of significantly longer training time (up to 4$\times$ on CPU to 7$\times$ on GPU) compared to other frameworks, and the experiment takes about 3 days to complete. By Table \ref{table:default-training-parameters-cifar10}, the default \#Epochs of TensorFlow on CIFAR-10 is 2560 with batch size of 128, which is significantly larger, and equivalent to \#Iterations $= (2560\times 50000)/128 = 1,000,000$ according to 
Formula~(\ref{formula:epoch-iteration}). 
We want to reduce the training time by cutting down the \#Epochs TensorFlow has set for CIFAR-10 while preserving its high accuracy. Interestingly, we found that reducing the \#Epochs by 10 times to 256 epochs, we can maintain very similar accuracy for CIFAR-10 on both CPU and GPU, as shown in Table \ref{table:tensorflow-2560-256-1}. Our empirical results are consistent to those reported by TensorFlow for 256 epochs~\cite{tensorflow} ($\sim$ 86\% accuracy). 
It is interesting to note that TensorFlow recommended default setting for \#Epochs is 10 times of 256, but its accuracy increase is only 0.30\% on CPU and 0.60\% on GPU. This translates to significantly longer training time, about 197,495.33s (54.86 hours) more on CPU and 11,117.83s (3.09 hours) more on GPU. Specifically, the training time of 2,560 epochs is 10.11$\times$ the training time of 256 epochs on CPU and 9.18$\times$ on GPU. On one hand, long training time spent once to train a model with higher prediction accuracy is well worth of the cost if real application deployment is highly sensitive to accuracy. On the other hand, some mission-critical applications may have training timing requirements, training for 256 epochs might be more attractive for two reasons: (1) the testing times of 2,560 epochs and 256 epochs are similar for both CPU and GPU respectively, and (2) the accuracy difference is only in the range of 0.30\% to 0.60\% for both CPU and GPU.   

\begin{table}[h!]
\centering
\caption{TensorFlow (TF) 2560 vs. 256 Epochs on CIFAR-10 (Server-1)}
\label{table:tensorflow-2560-256-1}
\vspace{-4mm}
\scalebox{0.8}{
\small
\begin{tabular}{|c|c|c|c|c|}
\hline
Framework & \#Epochs & \begin{tabular}[c]{@{}c@{}}Training\\ Time (s)\end{tabular} & \begin{tabular}[c]{@{}c@{}}Testing\\ Time (s)\end{tabular} & \begin{tabular}[c]{@{}c@{}}Accuracy\\ (\%)\end{tabular} \\ \hline
\multirow{2}{*}{ TF-CPU } & 2,560 & \DEFAULTHIGHLIGHT{219,169.14} & \DEFAULTHIGHLIGHT{4.80} & \DEFAULTHIGHLIGHT{86.90} \\
\cline{2-5}
& 256 & \textbf{21,673.81} & \textbf{4.79} & 86.60 \\
\hline
\multirow{2}{*}{ TF-GPU } & 2,560 & \DEFAULTHIGHLIGHT{12,477.05} & \DEFAULTHIGHLIGHT{2.34} & \DEFAULTHIGHLIGHT{87.00} \\
\cline{2-5}
& 256 & \textbf{1,359.22} & 2.44 & 86.40 \\
\hline
\end{tabular}
} 
\end{table}

\begin{table}[h!]
\vspace{-4mm}
\centering
\caption{Larger \#Iterations for Caffe (GPU-1)}
\label{table:caffe-improvement-epochs-gpu-1}
\vspace{-4mm}
\scalebox{0.8}{
\small
\begin{tabular}{|c|c|c|c|c|c|}
\hline
Dataset                   & \#Iterations & \#Epochs & \begin{tabular}[c]{@{}c@{}}Training\\ Time (s)\end{tabular} & \begin{tabular}[c]{@{}c@{}}Testing\\ Time (s)\end{tabular} & \begin{tabular}[c]{@{}c@{}}Accuracy\\ (\%)\end{tabular} \\ \hline
\multirow{4}{*}{MNIST}    & 10000        & 10.67    & \DEFAULTHIGHLIGHT{97.02}                                                       & \DEFAULTHIGHLIGHT{0.55}                                                       & \DEFAULTHIGHLIGHT{99.14$\pm$0.03}                                                   \\ \cline{2-6} 
                          & 15000        & 16.00       & 145.22                                                      & 0.55                                                       & 99.04                                                   \\ \cline{2-6} 
                          & 20000        & 21.33    & 194.20                                                       & \textbf{0.54}                                                      & {\bf 99.22}                                                   \\ \cline{2-6} 
                          & 240000       & 256.00      & 2,333.93                                                    & 0.55                                                       & 99.05                                                   \\ \hline
\multirow{4}{*}{CIFAR-10} & 5000         & 10.00       & \DEFAULTHIGHLIGHT{163.51}                                                      & \DEFAULTHIGHLIGHT{1.36}                                                       & \DEFAULTHIGHLIGHT{75.52}                                                   \\ \cline{2-6} 
                          & 500000       & 1000.00     & 16,305.99                                                    & 1.37                                                       & 76.90                                                    \\ \cline{2-6} 
                          & 1000000      & 2000.00     & 32,688.06                                                    & 1.43                                                       & 75.48                                                   \\ \cline{2-6} 
                          & 1280000      & 2560.00     & 41,945.34                                                    & 1.423                                                    & {\bf 77.34}                                                   \\ \hline
\end{tabular}
} 
\vspace{-5mm}
\end{table}

\begin{figure*}[h!]
\vspace{-3mm}
\centering
\subfloat[Varying \#Epochs (CPU)]{
  \centering
  \includegraphics[width=0.32\linewidth]{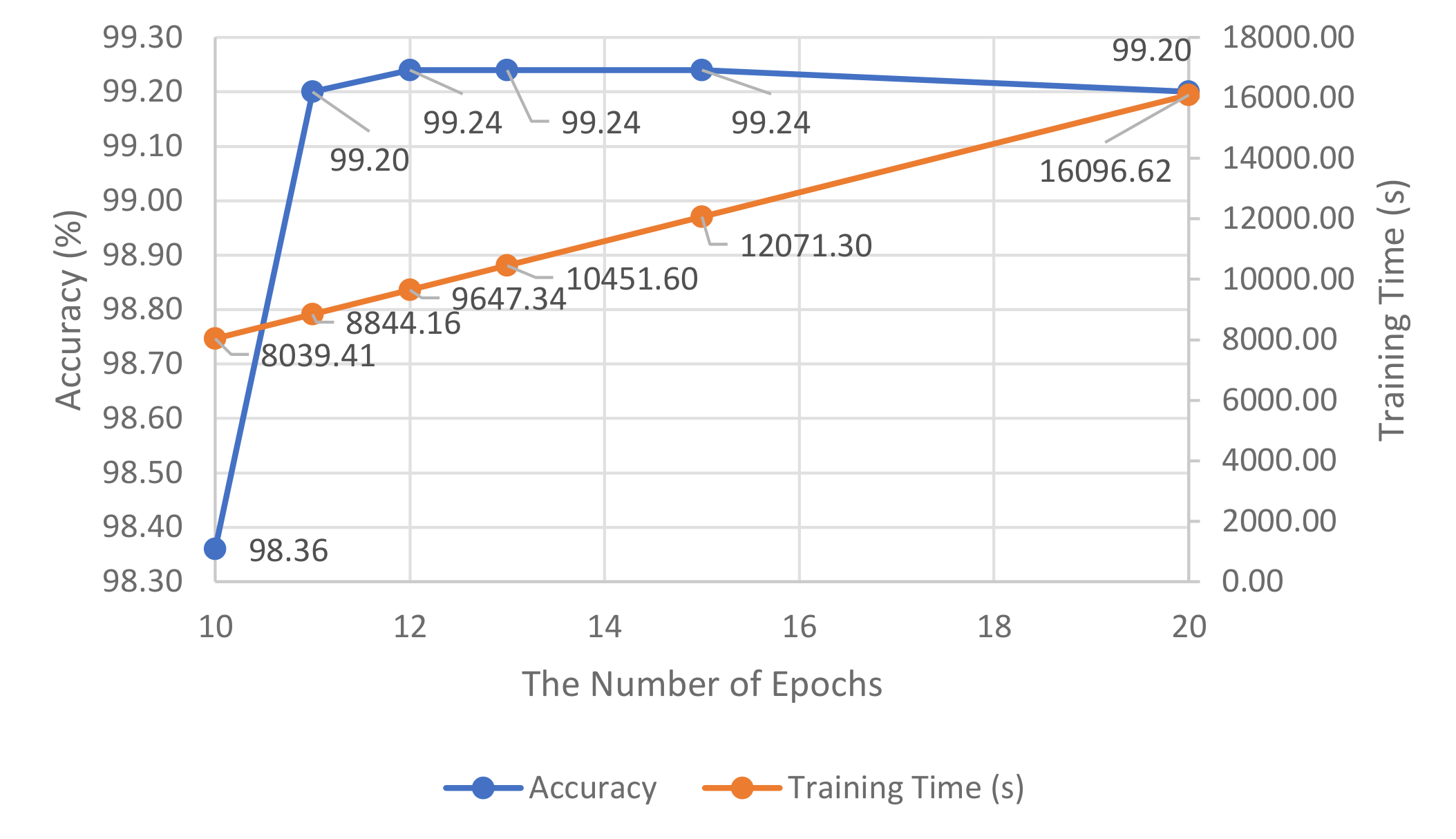}
  \label{fig:torch-improvement-epochs-mnist-cpu-1}
}
\subfloat[Varying \#Epochs (GPU)]{
  \centering
  \includegraphics[width=0.32\linewidth]{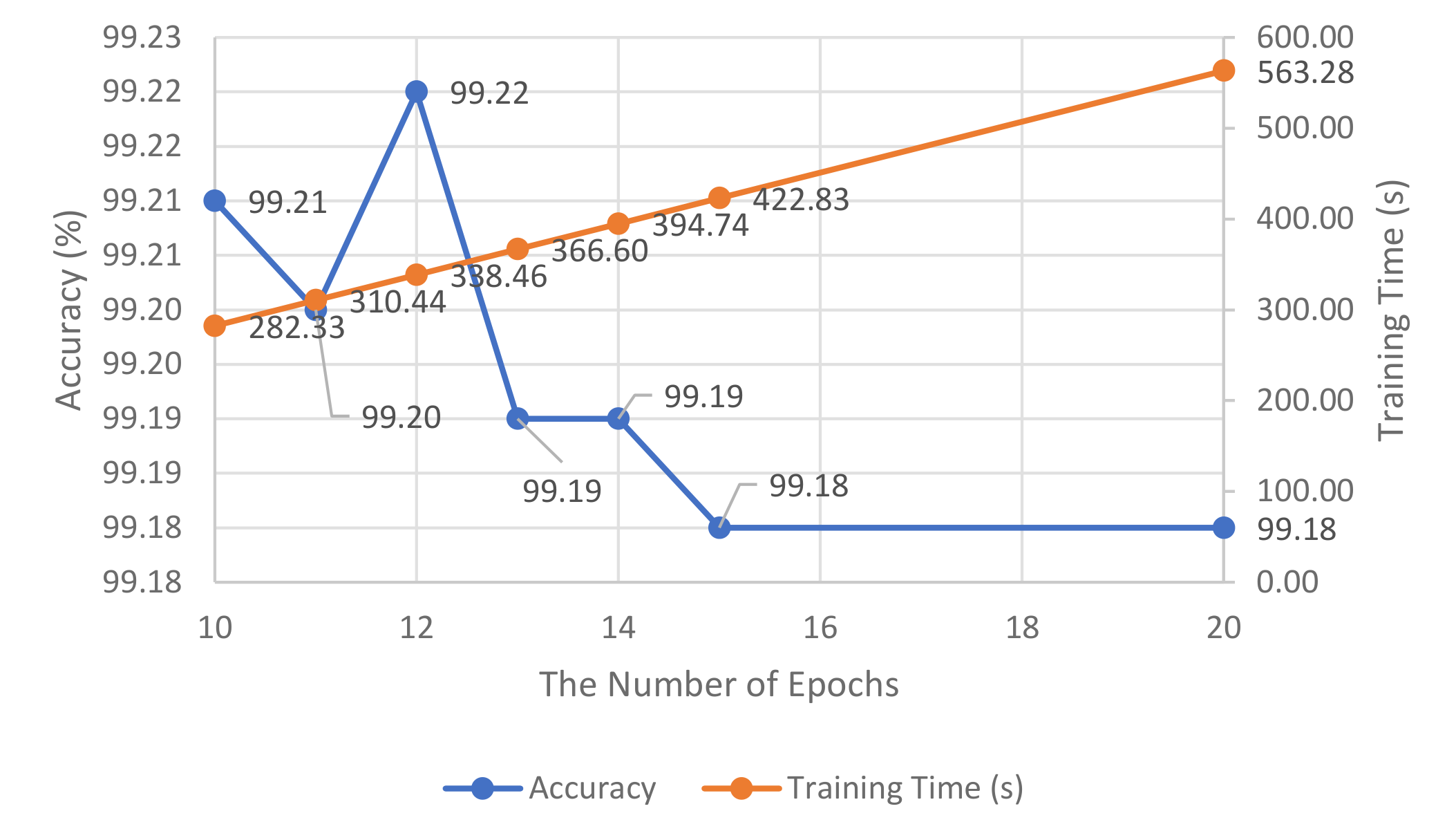}
  \label{fig:torch-improvement-epochs-mnist-gpu-1}
}
\subfloat[Varying Batch Size (GPU)]{
  \centering
  \includegraphics[width=0.32\textwidth]{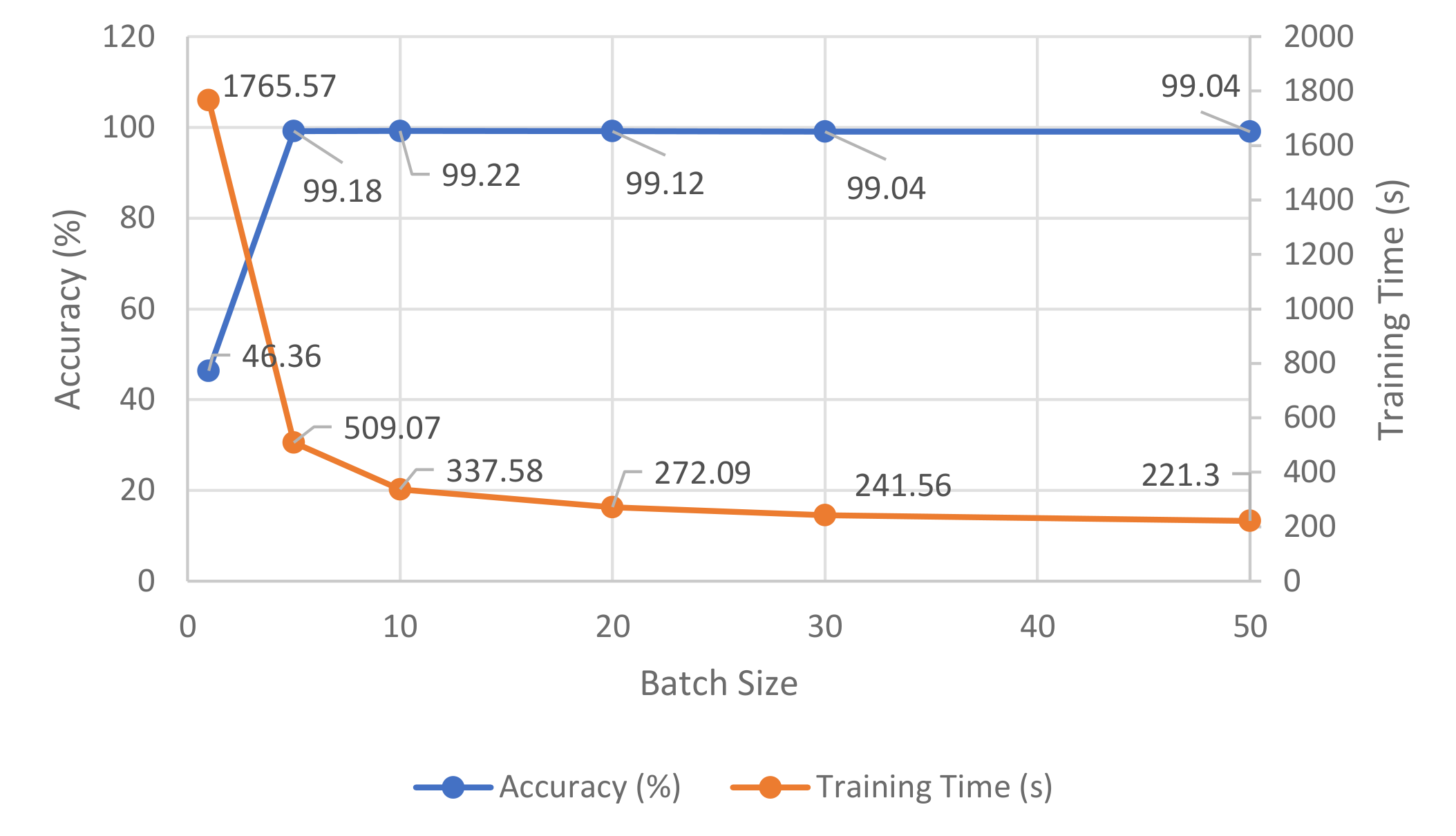}
  \label{fig:torch-improvement-bachsize-mnist-gpu-1}
} 
\caption{Experimental Results on MNIST, Accuracy and Training Time in Torch (Server-1)}
\label{fig:torch-improvement-mnist-1}
\vspace{-5mm}
\end{figure*}

\begin{figure*}[h!]
\centering
\subfloat[Varying \#Epochs (CPU)]{
  \centering
  \includegraphics[width=0.32\textwidth]{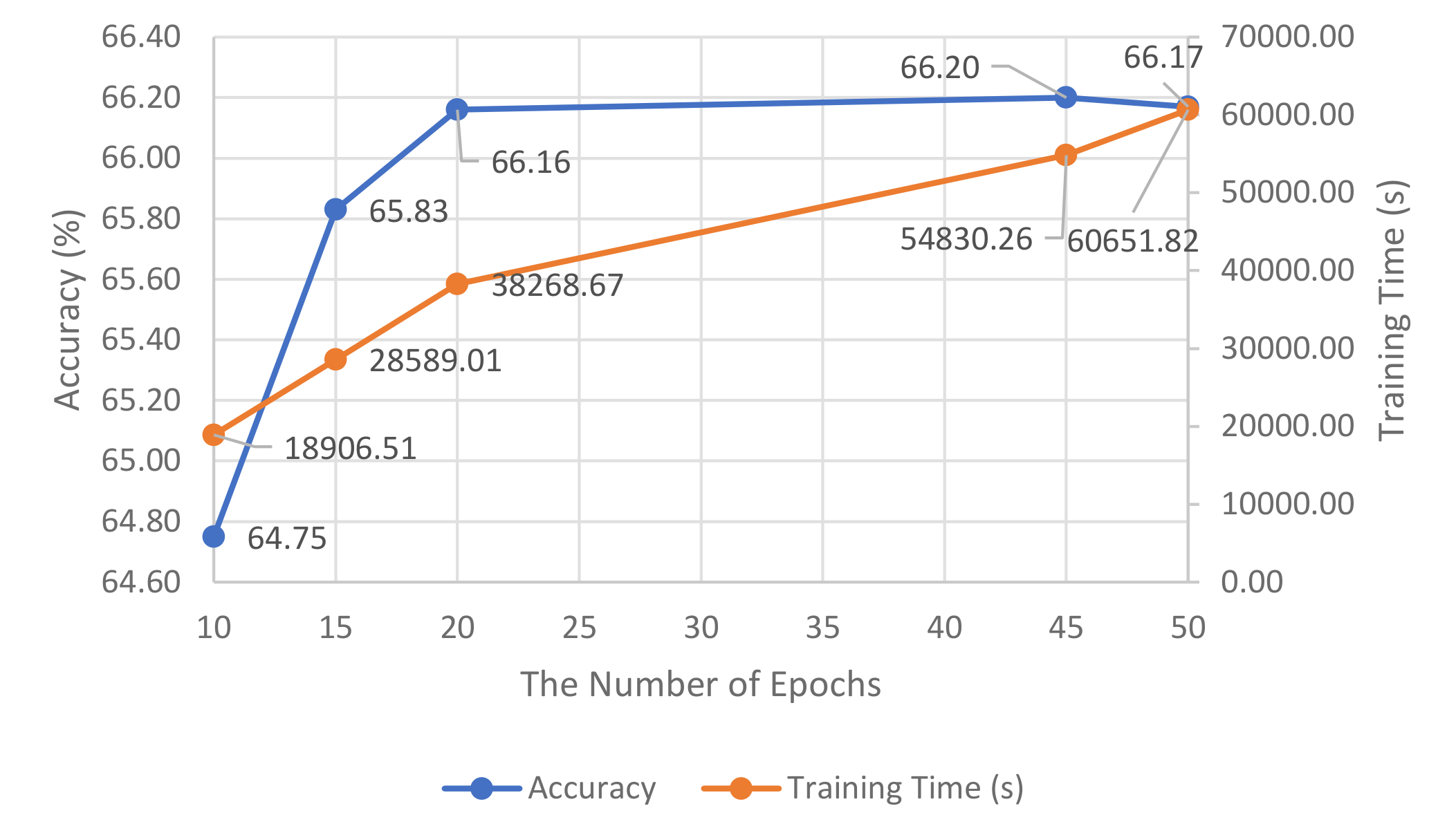}
  \label{fig:torch-improvement-epochs-cifar10-cpu-1}
} 
\subfloat[Varying \#Epochs (GPU)]{
  \centering
  \includegraphics[width=0.32\textwidth]{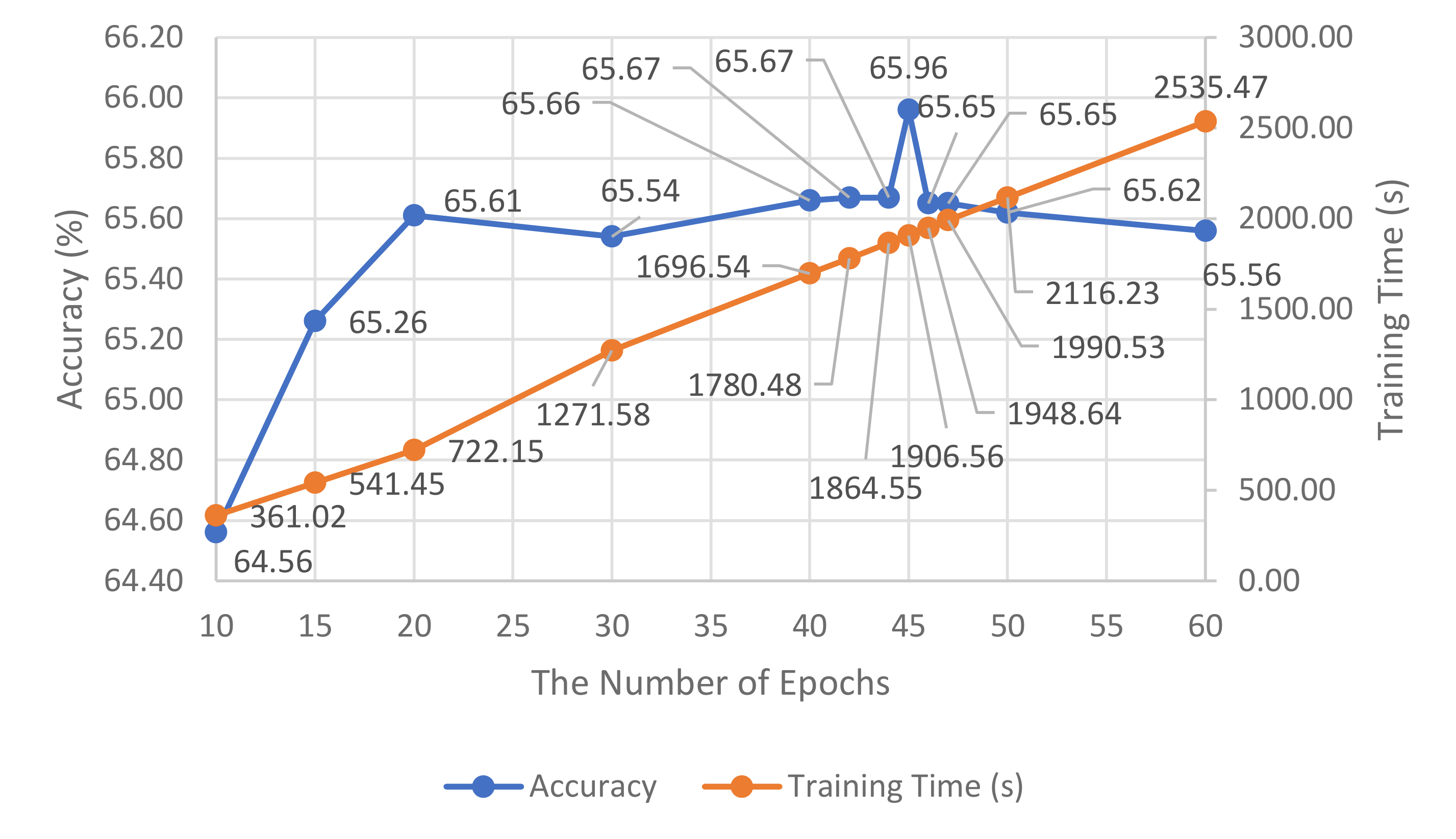}
  \label{fig:torch-improvement-epochs-cifar10-gpu-1}
} 
\subfloat[Varying Batch Size (GPU)]{
  \centering
  \includegraphics[width=0.32\textwidth]{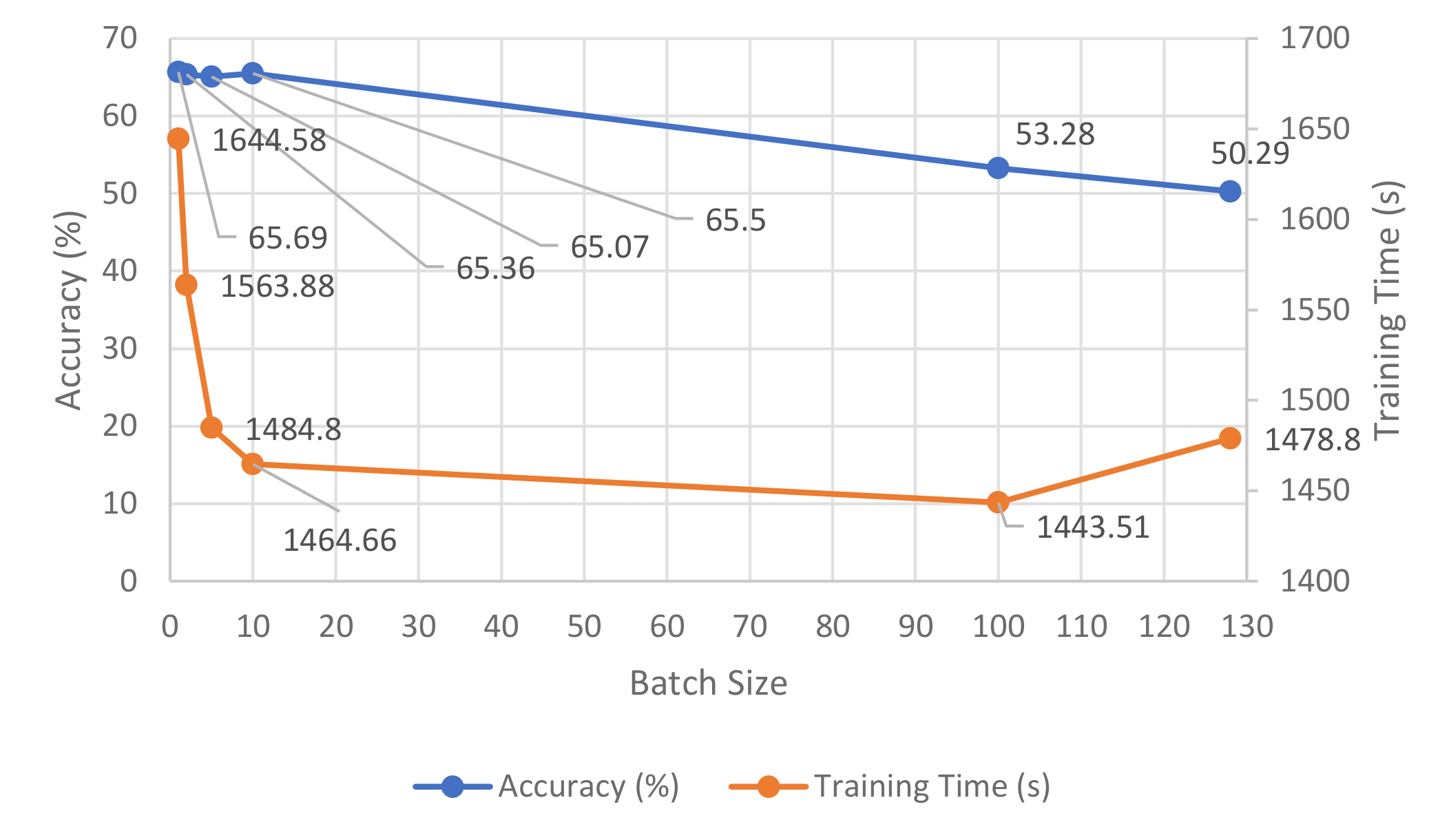}
  \label{fig:torch-improvement-batchsize-cifar10-gpu-1}
} 
\caption{Experimental Results on CIFAR-10,  Accuracy and Training Time in Torch (Server-1)}
\label{fig:torch-improvement-cifar10-1}
\vspace{-5mm}
\end{figure*}

\subsubsection{Caffe}
Caffe has shorter training time but the prediction accuracy of its trained DNN model is lower than TensorFlow and Torch on MNIST and lower than TensorFlow on CIFAR-10 for both CPU and GPU. 
From Table \ref{table:default-training-parameters-mnist} and \ref{table:default-training-parameters-cifar10}, it is observed that Caffe has the smallest default \#Iterations (\#Epochs) on both datasets. We next investigate the impact of larger \#Iterations on the accuracy of Caffe. Table~\ref{table:caffe-improvement-epochs-gpu-1} reports the measurement results.
We make two observations. 
{\em First}, 
for MNIST, training for 20,000 iterations (21.33 epochs) shows the higher accuracy of 99.22\%, which is 0.08\% higher than the Caffe's default (99.14\%), at an acceptable cost of additional 97.18s for training. 
It is also interesting to note that the accuracy of 15,000 iterations (16 epochs) and of 240,000 iterations (256 epochs) are lower than default by 0.10\% and 0.09\% respectively, even by adding longer training time of 48.2s and 2,236.91s respectively. One reason of the lower accuracy for the 15,000 iterations is probably that the training is trapped into a local optimum instead of the global one, and the lower accuracy of the 240,000 iterations is likely due to over-fitting, i.e., the trained model is over-fitted to the training dataset, losing its generalization on the testing dataset.
{\em Second}, 
for CIFAR-10, we use the default setting of \#Epochs in TensorFlow to cap the maximum \#Iterations to 1,280,000 (i.e., 2,560 epochs), which increases the accuracy of Caffe by 2.41\% at the cost of 255.53$\times$ more training time than the default.
Also, the 500,000 iterations help increase the prediction accuracy by 1.38\% at the cost of 99.72$\times$ training time, considerably lower than that of 1,280,000 iterations. However, careful examination of Table~\ref{table:caffe-improvement-epochs-gpu-1}, it also indicates that the correlation between the training time and the prediction accuracy is non-linear and more complex.
{\em In summary}, 
this set of experiments indicates that Caffe can benefit from larger \#Epochs (i.e., longer training time) in some cases, though the larger \#Epochs does not necessarily guarantee a higher accuracy. This is because local optimum and over-fitting may lower accuracy with more iterations. For content-rich dataset like CIFAR-10, larger \#Epochs helps improve the accuracy of Caffe. 

\subsubsection{Torch} 
The next set of experiments is designed to study the impact of hyper-parameter \#Epochs (\#Iterations) on the performance of Torch in terms of accuracy and training time. We keep the default for the DNN structure and other hyper-parameters and vary the \#Epochs on MNIST and CIFAR-10 for both CPU and GPU platforms. 

Figure \ref{fig:torch-improvement-mnist-1} (MNIST) and \ref{fig:torch-improvement-cifar10-1} (CIFAR-10) show the results and we use the left y-axis for accuracy and the right y-axis for training time to facilitate the analysis of the relationship between training time and accuracy. Overall, the training time increases as the \#Epochs increases in all cases. For accuracy, Figure \ref{fig:torch-improvement-epochs-mnist-cpu-1}, \ref{fig:torch-improvement-epochs-cifar10-cpu-1} show that the accuracy of Torch on CPU first increases rapidly to reach the optimal value and then stays stable or slightly drops probably due to over-fitting. For MNIST, the peak accuracy of 99.24\% is first achieved at the 12th epoch while the CPU training time is 9,647.34s. 
For CIFAR-10, the peak accuracy of 66.20\% is first obtained at the 45th epoch, with the CPU training time of 54,830.26s. However, Torch-CPU reaches the accuracy of 66.16\% at the 20th epoch, and as the \#Epoch increases from 20 to 45, the accuracy for Torch-CPU only increases slightly by 0.04\% from 66.16\% to 66.20\%. Compared to the peak accuracy at the 45th epoch, with only 0.04\% less than the peak accuracy, it can save the training time by 16,561.59s, approximately 30\% of the training time for the 45 epochs.

Figure \ref{fig:torch-improvement-epochs-mnist-gpu-1}, \ref{fig:torch-improvement-epochs-cifar10-gpu-1} show that Torch experiences more accuracy fluctuations on GPU for both MNIST and CIFAR-10, with the peak GPU accuracy of 99.22\% first at the 12th epoch for MNIST with the training time of 338.46s, and the peak GPU accuracy of 65.96\% first at the 45th epoch for CIFAR-10 with the training time of 1,906.56s.  Comparing Torch on CPU and GPU, the training time of GPU is about 28$\times$ faster than CPU with a small loss of accuracy. These experiments indicate that the accuracy and training time curve on CPU is smoother than that on GPU and they also illustrate why Torch has the default \#Epochs=12 for MNIST and \#Epochs=45 for CIFAR-10 (recall Table \ref{table:default-training-parameters-mnist} and \ref{table:default-training-parameters-cifar10}). For CIFAR-10, Torch also achieves 0.24\% higher accuracy on CPU than GPU, though the time spent for CPU training is 27.76$\times$ longer than its GPU counterpart. Thus, we show the impact of \#Epochs for Torch on accuracy and training time, and find its optimal \#Epochs on MNIST and CIFAR-10.

\begin{figure}[h!]
\vspace{-2mm}
\centering
\includegraphics[width=0.35\textwidth]{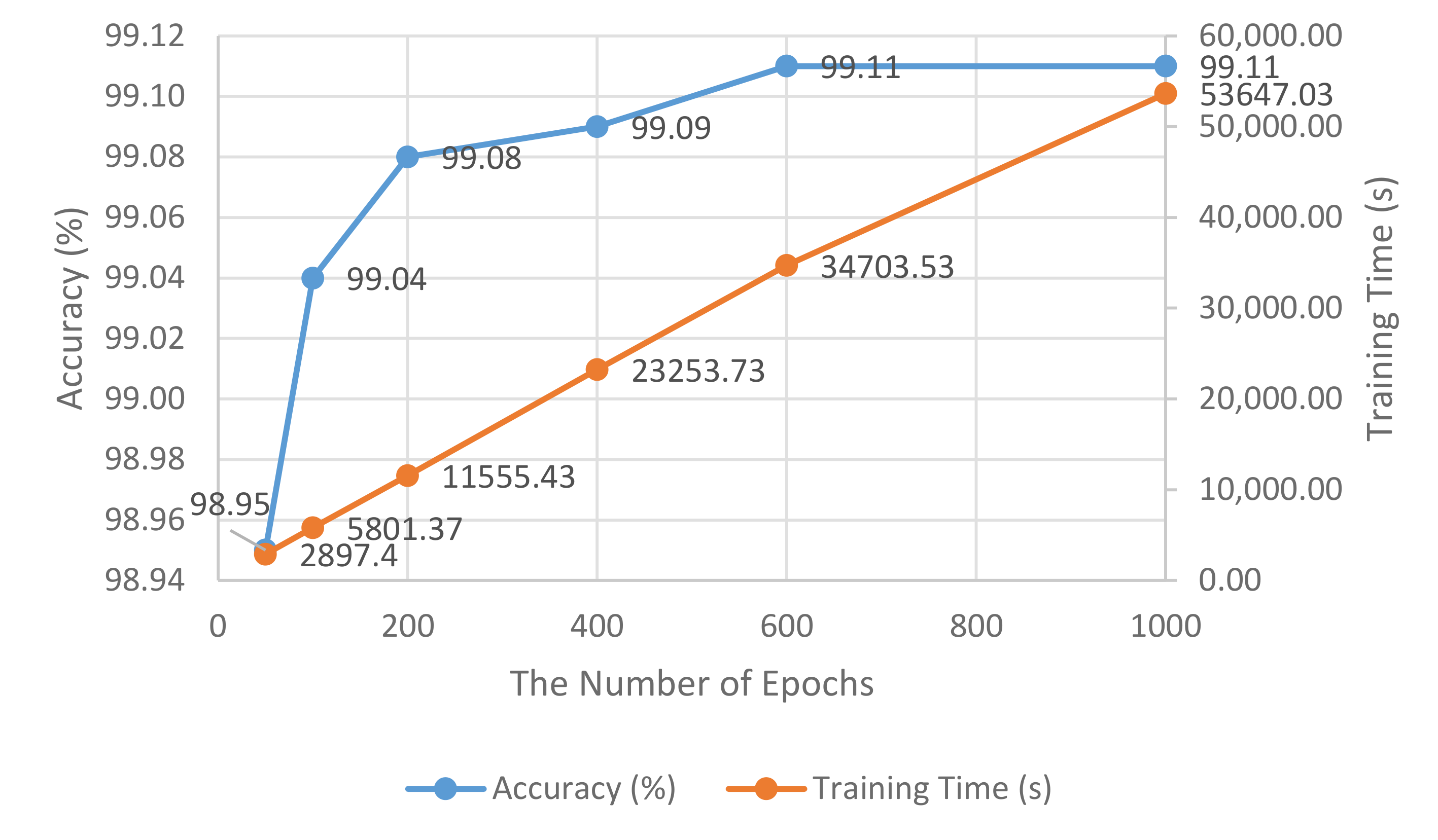}
\caption{Varying \#Epochs for Theano on MNIST (CPU)}
\label{fig:theano-improvement-mnist-cpu-1}
\vspace{-7mm}
\end{figure}

\subsubsection{Theano} \label{section:theano-improvement-epochs}

We report the experiments on the impact of varying \#Epochs on the performance of Theano for CPU in Figure \ref{fig:theano-improvement-mnist-cpu-1}. Both the training time and the accuracy increase as the \#Epochs increases, though the accuracy has a much slower response to the growth of \#Epochs. Theano first achieves the peak accuracy of 99.11\% at the 600th epoch, showing that Theano may benefit from larger \#Epochs to improve the accuracy.

In summary, all experiments conducted for studying the impact of \#Epochs consistently confirm two observations for all four frameworks: (1) The training time is proportional to \#Epochs independently of dataset or framework choices, and (2) a larger \#Epochs does not guarantee to increase the model accuracy but a peak accuracy threshold can be found.

\subsection{Impact of Hyper-parameter: Batch Size} 
\label{subsection:impact-batch-size}

The batch size is another important hyper-parameter for all DL frameworks. To understand the role of batch size on the performance of DL frameworks, we only vary the batch size in this set of experiments while keeping the DNN structure and the other hyper-parameters by their default.

\subsubsection{TensorFlow}

Table \ref{table:tensorflow-batchsize-impact-1} shows the training time (per epoch and total), testing time and the accuracy as the batch size increases. First, as the batch size starts to increase exponentially, the training time and the accuracy both start to drop. Since all other hyper-parameters remain their default configuration, one possible interpretation is that the larger batch size adds more complexity to the learning process and fewer features may be found per iteration. Moreover, a larger batch size implies the drop of the total \#Iterations (\#Iters) for a given default \#Epochs based on Formula~(\ref{formula:epoch-iteration}). However, when the batch size increases to $50,000$, the training time becomes higher than that of the default $batch\_size=50$, which is likely due to the use of virtual memory due to the large batch (see Section~\ref{section:cpu-memory} for more discussions). 

\begin{table}[h!]
\vspace{-4mm}
\centering
\caption{TensorFlow with Varying Batch Size on MNIST (CPU-1)}
\label{table:tensorflow-batchsize-impact-1}
\vspace{-4mm}
\scalebox{0.8}{
\small
\begin{tabular}{|c|c|c|c|c|c|c|}
\hline
\multirow{2}{*}{\begin{tabular}[c]{@{}c@{}}Batch\\ Size\end{tabular}} & \multirow{2}{*}{\#Iters} & \multirow{2}{*}{\#Epochs} & \multicolumn{2}{c|}{Training Time (s)} & \multirow{2}{*}{\begin{tabular}[c]{@{}c@{}}Testing\\ Time (s)\end{tabular}} & \multirow{2}{*}{\begin{tabular}[c]{@{}c@{}}Accuracy\\ (\%)\end{tabular}} \\ \cline{4-5}
                                                                      &                               &                           & Per Epoch           & Total            &                                                                             &                                                                          \\ \hline
50    & 20000 & 18.18 & \DEFAULTHIGHLIGHT{61.29} & \DEFAULTHIGHLIGHT{1,114.34} & \DEFAULTHIGHLIGHT{2.73} & \DEFAULTHIGHLIGHT{99.24$\pm$0.05} \\ \hline
500   & 2000  & 18.18 & 54.26 & 986.52   & 3.14 & 98.65 \\ \hline
5000  & 200   & 18.18 & \textbf{51.56} & \textbf{937.37}   & 3.16 & 95.89 \\ \hline
50000 & 20    & 18.18 & 64.88 & 1,179.68 & 3.34 & \textit{60.62} \\ \hline
\end{tabular}
}
\vspace{-3mm}
\end{table}

\subsubsection{Caffe}
We observe very similar behavior from Caffe. By Table \ref{table:caffe-batchsize-impact-1}, as the batch size increases, both the accuracy and the training time decrease, though the testing time remains almost unchanged. Caffe has the lowest accuracy (60.30\%) when the batch size increases to $60000$, indicating that the whole training process may not converge well.

\begin{table}[h!]
\vspace{-4mm}
\caption{Caffe with Varying Batch Size on MNIST (CPU-1)}
\label{table:caffe-batchsize-impact-1}
\vspace{-4mm}
\centering
\scalebox{0.8}{
\small
\begin{tabular}{|c|c|c|c|c|c|c|}
\hline
\multirow{2}{*}{\begin{tabular}[c]{@{}c@{}}Batch\\ Size\end{tabular}} & \multirow{2}{*}{\#Iters} & \multirow{2}{*}{\#Epochs} & \multicolumn{2}{c|}{Training Time (s)} & \multirow{2}{*}{\begin{tabular}[c]{@{}c@{}}Testing\\ Time (s)\end{tabular}} & \multirow{2}{*}{\begin{tabular}[c]{@{}c@{}}Accuracy\\ (\%)\end{tabular}} \\ \cline{4-5}
                                                                      &                               &                           & Per Epoch           & Total            &                                                                             &                                                                          \\ \hline

64    & 10000 & 10.67 & \DEFAULTHIGHLIGHT{48.02} & \DEFAULTHIGHLIGHT{512.18} & \DEFAULTHIGHLIGHT{3.33} & \DEFAULTHIGHLIGHT{99.04$\pm$0.02} \\ \hline
640   & 1000  & 10.67 & 44.83 & 478.22 & \textbf{3.14} & 98.70 \\ \hline
6400  & 100   & 10.67 & 39.79 & 424.38 & \textbf{3.14} & 93.66 \\ \hline
60000 & 10    & 10.00 & \textbf{37.72} & \textbf{377.16} & 3.16 & \textit{60.30} \\ \hline
\end{tabular}
}
\vspace{-3mm}
\end{table}

\subsubsection{Torch}

Recall Table \ref{table:default-training-parameters-mnist} and Table \ref{table:default-training-parameters-cifar10}, Torch has the smallest default batch sizes among all four DL frameworks. 
Figure \ref{fig:torch-improvement-bachsize-mnist-gpu-1} and \ref{fig:torch-improvement-batchsize-cifar10-gpu-1} show the results of varying the batch size for Torch for MNIST and CIFAR-10 respectively. For MNIST, we vary the batch size from 1 to 50. As the batch size increases from 1 to 5, the accuracy starts to increase quickly and reaches the peak accuracy of 99.22\% at the batch size of 10 and then starts to drop very slowly from 10 to 50, while the training time decreases at first quickly and then the trend becomes slow as the batch size increases beyond 10. The training time is 1,765.57s for $batch\_size = 1$, which is 3.47$\times$ that of $batch\_size = 5$ (509.07s). For CIFAR-10, Figure~\ref{fig:torch-improvement-batchsize-cifar10-gpu-1} shows that the batch size of 1 is optimal and as soon as the batch size starts to increase, the prediction accuracy starts to drop while the training time drops sharply when the batch size increases from 1 to 10 and then drops at much slower pace until the batch size reaches 100. This set of experiments confirms that Torch achieves the highest accuracy at $batch\_size = 10$ for MNIST and $batch\_size = 1$ for CIFAR-10. We conclude with three interesting observations: (1) For CIFAR-10, the accuracy of $batch\_size = 10$ is 65.50\%, only 0.19\% lower than the optimal $batch\_size=1$ but Torch at $batch\_size = 10$ enjoys a training time reduction of 179.92s, about 11\% of the training time for $batch\_size = 1$. (2) Torch shows the worst accuracy with $batch\_size=1$ on MNIST while it achieves the best accuracy on CIFAR-10, showing that its default configuration is highly dataset dependent. This is because when the batch size is 1, the training is pure stochastic, implying the features extracted from each batch is partial. Moreover, the LR on MNIST is 0.1 much larger than that for CIFAR-10 (0.01). The higher LR may lead to over-fitting more easily on partial features due to the low entropy of a dataset, e.g., MNIST. (3) Table \ref{table:torch-batchsize-impact-1} shows that the experimental results for much larger batch sizes. When the batch size increases from 10 to 1000, the accuracy drops slightly and the training time drops to 19.21\%. However, when $batch\_size=10000$, the training time increases compared to that for $batch\_size=1000$. One reason could be the higher memory overhead. When we further increase the batch size to $60000$, Torch crashed on both platforms.

\begin{table}[h!]
\vspace{-4mm}
\centering
\caption{Torch with Varying Batch Size on MNIST (CPU-1)}
\label{table:torch-batchsize-impact-1}
\vspace{-4mm}
\scalebox{0.8}{
\small
\begin{tabular}{|c|c|c|c|c|c|c|}
\hline
\multirow{2}{*}{\begin{tabular}[c]{@{}c@{}}Batch\\ Size\end{tabular}} & \multirow{2}{*}{\#Iters} & \multirow{2}{*}{\#Epochs} & \multicolumn{2}{c|}{Training Time (s)} & \multirow{2}{*}{\begin{tabular}[c]{@{}c@{}}Testing\\ Time (s)\end{tabular}} & \multirow{2}{*}{\begin{tabular}[c]{@{}c@{}}Accuracy\\ (\%)\end{tabular}} \\ \cline{4-5}
                                                                      &                               &                           & Per Epoch           & Total            &                                                                             &                                                                          \\ \hline
10    & 72000 & 12 & \DEFAULTHIGHLIGHT{803.95} & \DEFAULTHIGHLIGHT{9647.34} & \DEFAULTHIGHLIGHT{56.52} & \DEFAULTHIGHLIGHT{99.24$\pm$0.00} \\ \hline
100   & 7200  & 12 & 256.86 & 3082.26 & 21.92 & 98.86 \\ \hline
1000  & 720   & 12 & \textbf{154.45} & \textbf{1853.41} & \textbf{14.16} & 96.89 \\ \hline
10000 & 72    & 12 & 194.73 & 2336.76 & 18.47 & \textit{83.61} \\ \hline
\end{tabular}
}
\vspace{-3mm}
\end{table}


\subsubsection{Theano}
Table \ref{table:theano-batchsize-impact-1} shows the measurement of Theano when batch size is varied from 50, 500 to 5000. Theano adopts the early stopping technique to address over-fitting~\cite{theano}. Thus, when $batch\_size=50$, it stopped earlier at the 178th epoch with the highest accuracy. The training time per epoch drops when the batch size increases from 500 to 5000 but increases when the batch size increases from 50 to 500. As the batch size increases, the accuracy declines. Also, Theano produces the same accuracy on both single GPU and multi-GPU platforms for the same batch size settings, demonstrating its good stability of accuracy. 

\begin{table}[h!]
\vspace{-4mm}
\centering
\caption{Theano with Varying Batch Size on MNIST (CPU-1)}
\label{table:theano-batchsize-impact-1}
\vspace{-4mm}
\scalebox{0.8}{
\small
\begin{tabular}{|c|c|c|c|c|c|c|}
\hline
\multirow{2}{*}{\begin{tabular}[c]{@{}c@{}}Batch\\ Size\end{tabular}} & \multirow{2}{*}{\#Iters} & \multirow{2}{*}{\#Epochs} & \multicolumn{2}{c|}{Training Time (s)} & \multirow{2}{*}{\begin{tabular}[c]{@{}c@{}}Testing\\ Time (s)\end{tabular}} & \multirow{2}{*}{\begin{tabular}[c]{@{}c@{}}Accuracy\\ (\%)\end{tabular}} \\ \cline{4-5}
                                                                      &                               &                           & Per Epoch           & Total            &                                                                             &                                                                          \\ \hline
50   & 178000 & 178 & 53.90 & \textbf{9594.97}  & 4.74 & \textbf{99.17} \\ \hline
500  & 20000  & 200 & \DEFAULTHIGHLIGHT{57.78} & \DEFAULTHIGHLIGHT{11555.43} & \DEFAULTHIGHLIGHT{4.49} & \DEFAULTHIGHLIGHT{99.08$\pm$0.00} \\ \hline
5000 & 2000   & 200 & \textbf{51.20} & 10239.11 & \textbf{4.01} & 98.61 \\ \hline
\end{tabular}
}
\vspace{-3mm}
\end{table}

\subsection{Impact of Tuning Multiple Hyper-parameters}

We have studied the impact of single hyper-parameter, such as \#Epochs and the batch size on the performance of DL frameworks, by varying one hyper-parameter while keeping the default setting for the rest of the parameters (recall Section~\ref{section:baseline-experiments}). Though larger \#Epochs may improve the accuracy at the cost of training time, and a larger batch size may decrease the accuracy, the correlation of \#Epochs and the batch size is much more complex. We dedicate the next set of experiments to study the impact of tuning multiple hyper-parameters and to understand whether such tuning may improve the accuracy for DL frameworks. 

\subsubsection{TensorFlow}
Based on the empirical analysis for the single hyper-parameter \#Iterations (\#Epochs) and batch size, we choose the $batch\_size = 500$ and $5000$ and the \#Iterations of 200, 2000 and 20000 to study the impact of tuning multiple hyper-parameters. 
Table \ref{table:tensorflow-batchsize-iterations} shows that $batch\_size = 500$ with 20000 iterations achieved much better accuracy than $batch\_size = 500$ with 2000 iterations (98.65\%). Also, by tuning the two hyper-parameters together, when $batch\_size = 5000$ and $\#Iterations=20000$, TensorFlow achieves the highest accuracy of 99.34\%, better than the accuracy (99.24\%) of its default configuration for MNIST, though at a high cost due to about 84.31$\times$ longer training time. Overall, this set of experiments indicates a larger batch size needs much longer training to converge and achieve higher accuracy, and thus, a fixed \#Epochs is not suitable for larger batch sizes.

\begin{table}[h!]
\vspace{-4mm}
\centering
\caption{TF with Varying Batch Size \& \#Iterations on MNIST (CPU-1)}
\label{table:tensorflow-batchsize-iterations}
\vspace{-4mm}
\scalebox{0.8}{
\small
\begin{tabular}{|c|c|c|c|c|}
\hline
\begin{tabular}[c]{@{}c@{}}Batch\\ Size\end{tabular} & \#Iterations & \begin{tabular}[c]{@{}c@{}}Training\\ Time (s)\end{tabular} & \begin{tabular}[c]{@{}c@{}}Testing\\ Time (s)\end{tabular} & \begin{tabular}[c]{@{}c@{}}Accuracy\\ (\%)\end{tabular} \\ \hline
50                                                   & 20000        & \DEFAULTHIGHLIGHT{1114.34}                                                     & \DEFAULTHIGHLIGHT{2.73}                                                       & \DEFAULTHIGHLIGHT{99.24$\pm$0.05}                                                   \\ \hline
500                                                  & 20000        & 9593.96                                                     & 3.14                                                       & 99.28                                                   \\ \hline
500                                                  & 2000         & 986.52                                                      & 3.14                                                       & 98.65                                                   \\ \hline
5000                                                 & 20000        & 93946.05                                                    & 3.17                                                       & \textbf{99.34}                                                   \\ \hline
5000                                                 & 200          & \textbf{937.37}                                                     & 3.16                                                       & 95.89                                                   \\ \hline
\end{tabular}
} 
\vspace{-3mm}
\end{table}

\subsubsection{Caffe}
Similarly, we choose $batch\_size = 640$ and $6400$ for Caffe. Table \ref{table:caffe-batchsize-iterations} shows that using $batch\_size = 640$ and $6400$ with 10000 iterations, Caffe can improve the accuracy (99.04\%) of its default configuration and achieve higher accuracy of 99.09\% at the cost of much longer training time (i.e., 4,199.61s for $batch\_size = 640$ compared to 512.18s for the default batch size of 64). This also indicates that a larger batch size may help increase the resistance to over-fitting.

\begin{table}[h!]
\vspace{-4mm}
\centering
\caption{Caffe with Varying Batch Size \& \#Iterations on MNIST (CPU-1)}
\label{table:caffe-batchsize-iterations}
\vspace{-4mm}
\scalebox{0.8}{
\small
\begin{tabular}{|c|c|c|c|c|}
\hline
\begin{tabular}[c]{@{}c@{}}Batch\\ Size\end{tabular} & \#Iterations & \begin{tabular}[c]{@{}c@{}}Training\\ Time (s)\end{tabular} & \begin{tabular}[c]{@{}c@{}}Testing\\ Time (s)\end{tabular} & \begin{tabular}[c]{@{}c@{}}Accuracy\\ (\%)\end{tabular} \\ \hline
64         & 10000        & \DEFAULTHIGHLIGHT{512.18}            & \DEFAULTHIGHLIGHT{3.33}            & \DEFAULTHIGHLIGHT{99.04$\pm$0.02}         \\ \hline
640        & 10000        & 4199.61           & \textbf{3.14}        & {\bf 99.09}         \\ \hline
640        & 1000         & 478.22            & \textbf{3.14}            & 98.70          \\ \hline
6400       & 10000        & 36408.52          & 3.16            & 99.05         \\ \hline
6400       & 100          & \textbf{424.38}            & \textbf{3.14}            & 93.66         \\ \hline
\end{tabular}
} 
\vspace{-3mm}
\end{table}

\subsubsection{Torch}

In this set of experiments, similar to Caffe and TensorFlow, $batch\_size = 100$ is chosen. Torch uses \#Epochs to control the training process. 
The default $\#Epochs$ is 12 and the default $batch\_size$ is 10 in Torch for MNIST.
Figure~\ref{fig:torch-accuracy-improvement-gpu-1} shows the accuracy measurements by varying the setting of \#Epochs with $batch\_size = 100$. With the larger batch size, as the \#Epochs increases, Torch achieves much better accuracy. Particularly, when $batch\_size = 100$, combined with $\#Epochs=80$ or $\#Epochs=120$, Torch achieved an higher accuracy of 99.31\% or 99.32\% respectively, compared with 99.22\% using its default $\#Epochs=12$ for MNIST (recall Table \ref{table:default-training-parameters-mnist} and Table \ref{table:default-server-1}). 
Overall, this set of experiments indicates that with larger \#Epochs and a larger batch size, Torch can improve the accuracy. Also a larger batch size tends to be more resilient to over-fitting.

\subsubsection{Theano}

\begin{figure*}[h!]
\vspace{-3mm}
\centering
\subfloat[TensorFlow]{
  \centering
  \includegraphics[width=0.24\linewidth]{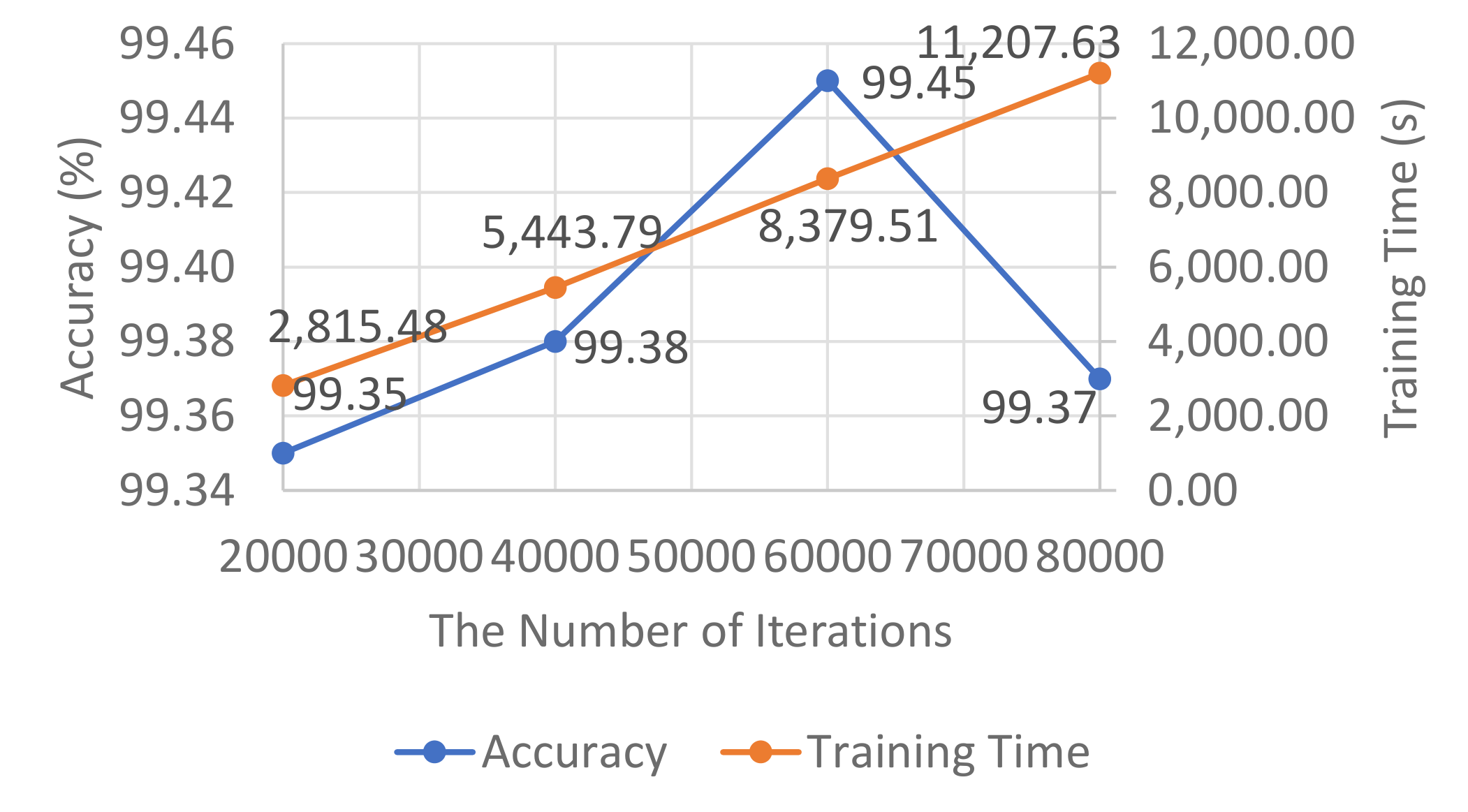}
  \label{fig:tensorflow-accuracy-improvement-gpu-1}
}
\subfloat[Caffe]{
  \centering
  \includegraphics[width=0.24\linewidth]{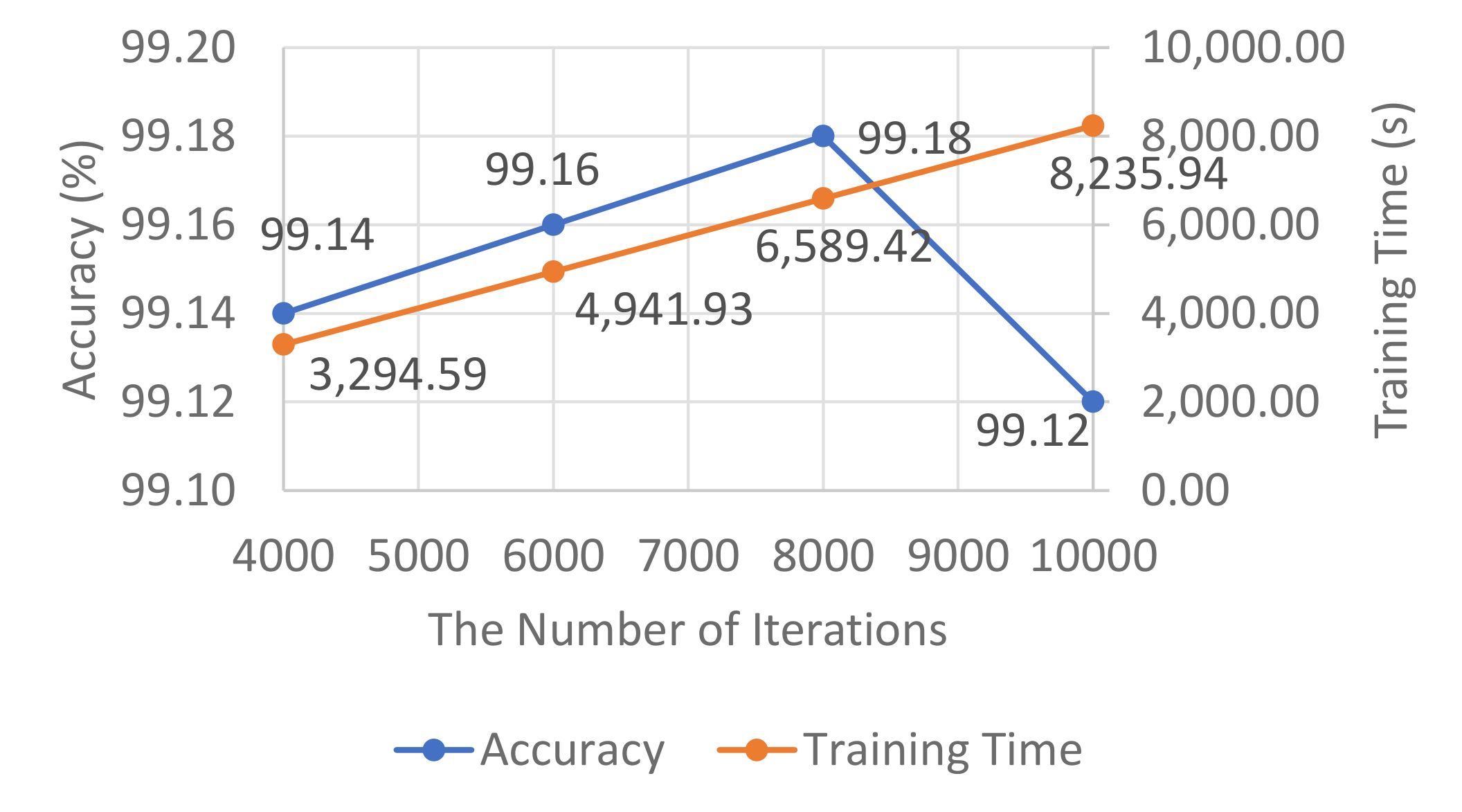}
  \label{fig:caffe-accuracy-improvement-gpu-1}
}
\subfloat[Torch]{
  \centering
  \includegraphics[width=0.24\linewidth]{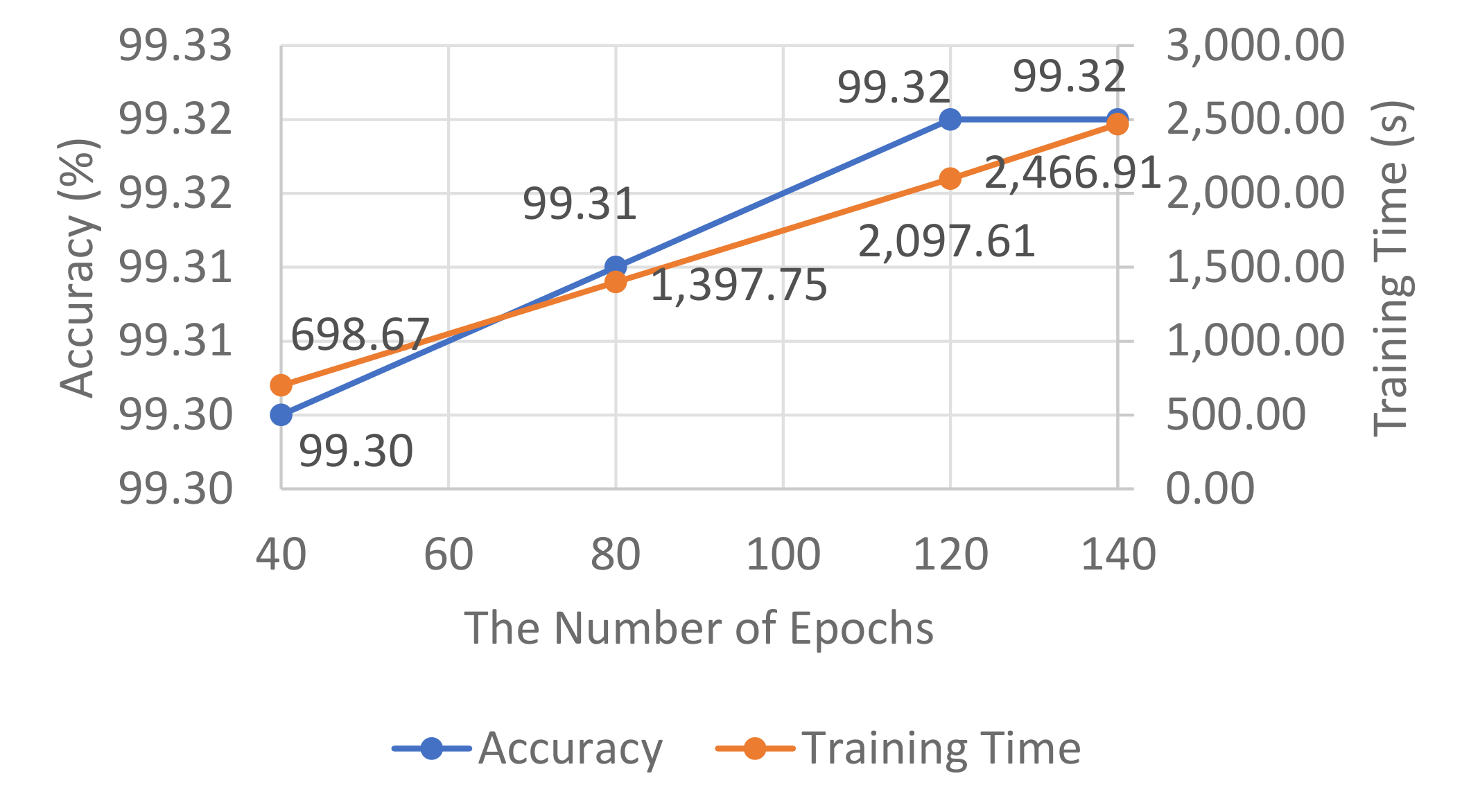}
  \label{fig:torch-accuracy-improvement-gpu-1}
}
\subfloat[Theano]{
  \centering
  \includegraphics[width=0.24\linewidth]{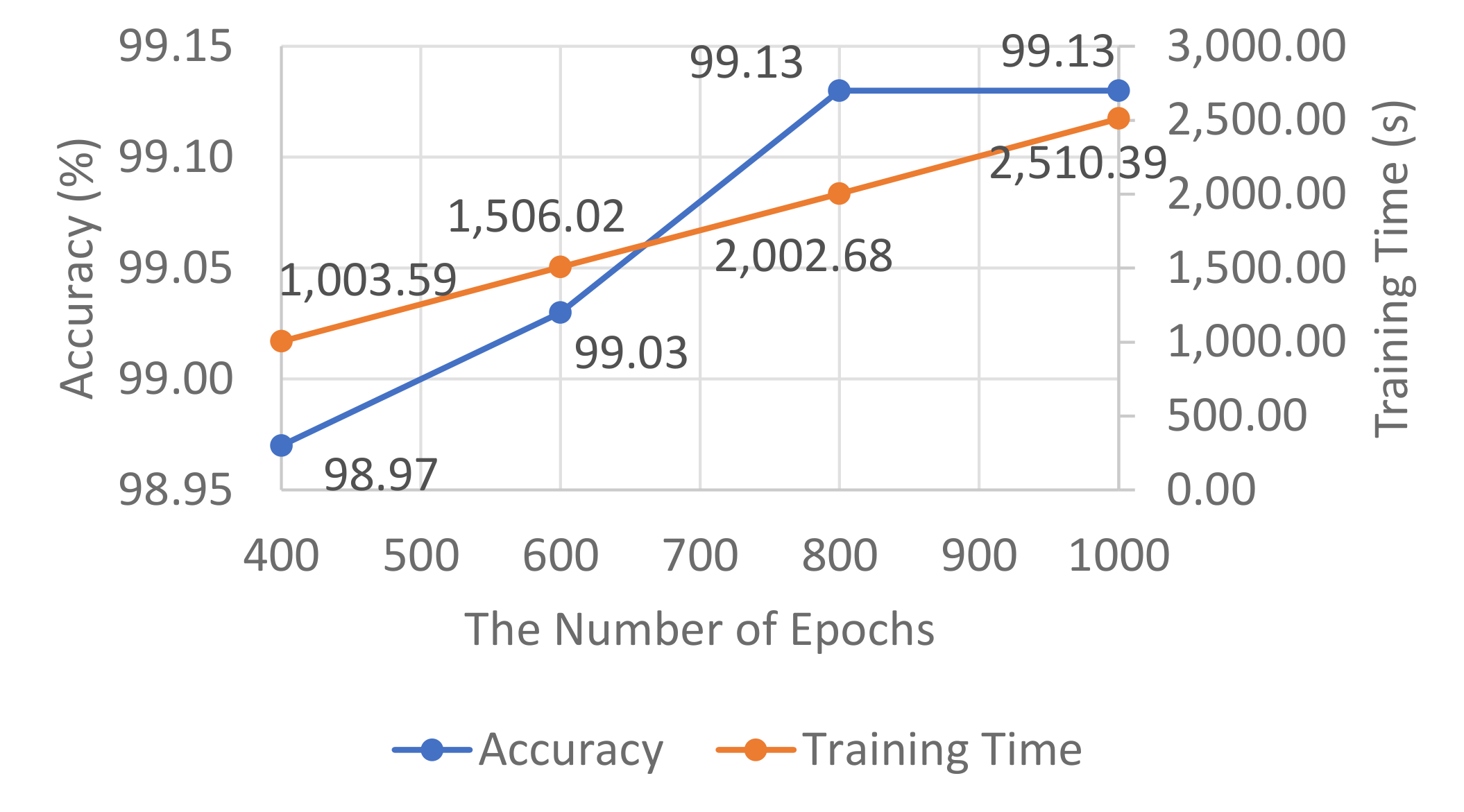}
  \label{fig:theano-accuracy-improvement-gpu-1}
}
\caption{Experimental Results on MNIST, Accuracy and Training Time with Varying \#Iterations(\#Epochs) (GPU)}
\label{fig:accuracy-improvement-mnist-1}
\vspace{-4mm}
\end{figure*}

As for Theano, the default $batch\_size = 500$ and the default $\#Epochs=200$ (recall Table \ref{table:default-training-parameters-mnist}). In this set of experiments, we set $batch\_size = 5000$ and vary the \#Epochs up to 1000. The experimental results are shown in Figure~\ref{fig:theano-accuracy-improvement-gpu-1}.
From 200 to 1000 epochs, the accuracy for $batch\_size = 5000$ increases continuously. From 800 epochs to 1000 epochs, Theano achieves the highest accuracy of 99.13\%, compared to the accuracy of 99.05\% when using its default setting of 200 epochs (recall Table \ref{table:default-server-1-mnist} Server-1: Theano-GPU for MNIST).

\begin{table*}[h!]
\centering
\caption{Accuracy Improvement on MNIST (GPU-1)}
\label{table:accuracy-improvement-mnist}
\vspace{-4mm}
\scalebox{0.8}{
\small
\begin{tabular}{|c|c|c|c|c|c|c|c|c|}
\hline
\multirow{2}{*}{Framework}  & \multirow{2}{*}{\begin{tabular}[c]{@{}c@{}}Batch\\ Size\end{tabular}} & \multirow{2}{*}{\#Iterations} & \multirow{2}{*}{\#Epochs} & \multirow{2}{*}{\begin{tabular}[c]{@{}c@{}}Learning\\ Rate\end{tabular}} & \multicolumn{2}{c|}{Training Time (s)} & \multirow{2}{*}{\begin{tabular}[c]{@{}c@{}}Testing\\ Time (s)\end{tabular}} & \multirow{2}{*}{\begin{tabular}[c]{@{}c@{}}Accuracy\\ (\%)\end{tabular}} \\ \cline{6-7}
                            &                                                                       &                               &                           &                                                                          & Per Epoch          & Total             &                                                                             &                                                                          \\ \hline
\multirow{9}{*}{TF} & \multirow{4}{*}{50}                                                   & 20000                         & 18.18                     & \multirow{2}{*}{0.0001}                                                  & \DEFAULTHIGHLIGHT{3.77}               & \DEFAULTHIGHLIGHT{68.51}             & \DEFAULTHIGHLIGHT{0.26}                                                                        & \DEFAULTHIGHLIGHT{99.21$\pm$0.03}                                                                    \\ \cline{3-4} \cline{6-9} 
                            &                                                                       & 60000                         & 54.55                     &                                                                          & 5.89               & 321.10            & 0.28                                                                        & \textbf{99.38}                                                                    \\ \cline{3-9} 
                            &                                                                       & 20000                         & 18.18                     & \multirow{2}{*}{0.001}                                                   & 5.94               & 107.91            & 0.28                                                                        & \textbf{99.26}                                                                    \\ \cline{3-4} \cline{6-9} 
                            &                                                                       & 60000                         & 54.55                     &                                                                          & 5.87               & 320.29            & 0.27                                                                        & 99.11                                                                    \\ \cline{2-9} 
                            & \multirow{5}{*}{5000}                                                 & 20000                         & 1,818.18                  & \multirow{2}{*}{0.0001}                                                  & 1.55               & 2,819.68          & 0.28                                                                        & \textbf{99.24}                                                                    \\ \cline{3-4} \cline{6-9} 
                            &                                                                       & 60000                         & 5,454.55                  &                                                                          & 1.56               & 8,484.47          & 0.28                                                                        & \textbf{99.37}                                                                    \\ \cline{3-9} 
                            &                                                                       & 20000                         & 1,818.18                  & \multirow{2}{*}{0.001}                                                   & 1.55               & 2,815.48          & 0.28                                                                        & \textbf{99.35}                                                                    \\ \cline{3-4} \cline{6-9} 
                            &                                                                       & \textbf{60000}                         & \textbf{5,454.55}                  &                                                                          & 1.54               & 8,379.51          & 0.28                                                                        & \textbf{99.45}                                                                    \\ \cline{3-9} 
                            &                                                                       & 60000                         & 5,454.55                  & 0.01                                                                     & \textbf{1.53}               & 8,331.56          & 0.28                                                                        & 98.59                                                                    \\ \hline
\multirow{9}{*}{Caffe}      & \multirow{4}{*}{64}                                                   & 8000                          & 8.53                      & \multirow{2}{*}{0.01}                                                    & 9.16               & \textbf{78.16}             & 0.55                                                                        & 99.02                                                                    \\ \cline{3-4} \cline{6-9} 
                            &                                                                       & 10000                         & 10.67                     &                                                                          & \DEFAULTHIGHLIGHT{9.10}               & \DEFAULTHIGHLIGHT{97.02}             & \DEFAULTHIGHLIGHT{0.55}                                                                        & \DEFAULTHIGHLIGHT{99.14$\pm$0.03}                                                                    \\ \cline{3-9} 
                            &                                                                       & 8000                          & 8.53                      & \multirow{2}{*}{0.1}                                                     & 9.30               & 79.32             & 0.56                                                                        & NaN                                                                      \\ \cline{3-4} \cline{6-9} 
                            &                                                                       & 10000                         & 10.67                     &                                                                          & 9.24               & 98.53             & 0.56                                                                        & NaN                                                                      \\ \cline{2-9} 
                            & \multirow{5}{*}{6400}                                                 & 8000                          & 853.33                    & \multirow{2}{*}{0.01}                                                    & 7.72               & 6,591.53          & 0.55                                                                        & 99.00                                                                    \\ \cline{3-4} \cline{6-9} 
                            &                                                                       & 10000                         & 1,066.67                  &                                                                          & 7.72               & 8,238.17          & 0.56                                                                        & 99.01                                                                    \\ \cline{3-9} 
                            &                                                                       & \textbf{8000}                          & \textbf{853.33}                    & \multirow{2}{*}{0.1}                                                     & 7.72               & 6,589.42          & 0.57                                                                        & \textbf{99.18}                                                                    \\ \cline{3-4} \cline{6-9} 
                            &                                                                       & 10000                         & 1,066.67                  &                                                                          & 7.72               & 8,235.94          & 0.56                                                                        & 99.12                                                                    \\ \cline{3-9} 
                            &                                                                       & 8000                          & 853.33                    & 0.5                                                                      & \textbf{7.66}               & 6,536.47          & 0.55                                                                        & NaN                                                                      \\ \hline
\multirow{9}{*}{Torch}      & \multirow{4}{*}{10}                                                   & 72000                         & 12                        & \multirow{2}{*}{0.05}                                                    & \DEFAULTHIGHLIGHT{28.21}              & \DEFAULTHIGHLIGHT{338.46}            & \DEFAULTHIGHLIGHT{1.73}                                                                        & \DEFAULTHIGHLIGHT{99.22$\pm$0.00}                                                                    \\ \cline{3-4} \cline{6-9} 
                            &                                                                       & 720000                        & 120                       &                                                                          & 28.29              & 3,394.76          & 1.79                                                                        & 99.18                                                                    \\ \cline{3-9} 
                            &                                                                       & 72000                         & 12                        & \multirow{2}{*}{0.4}                                                     & 28.50              & 341.99            & 1.81                                                                        & 36.43                                                                    \\ \cline{3-4} \cline{6-9} 
                            &                                                                       & 720000                        & 120                       &                                                                          & 29.18              & 3,501.25          & 1.79                                                                        & 65.57                                                                    \\ \cline{2-9} 
                            & \multirow{5}{*}{100}                                                  & 7200                          & 12                        & \multirow{2}{*}{0.05}                                                    & 17.53              & 210.36            & 1.64                                                                        & 98.88                                                                    \\ \cline{3-4} \cline{6-9} 
                            &                                                                       & 72000                         & 120                       &                                                                          & 17.54              & 2,105.31          & 1.64                                                                        & 99.04                                                                    \\ \cline{3-9} 
                            &                                                                       & 7200                          & 12                        & \multirow{2}{*}{0.4}                                                     & \textbf{17.47}              & \textbf{209.64}            & \textbf{1.62}                                                                        & \textbf{99.24}                                                                    \\ \cline{3-4} \cline{6-9} 
                            &                                                                       & \textbf{72000}                         & \textbf{120}                       &                                                                          & 17.48              & 2,097.61          & \textbf{1.62}                                                                        & \textbf{99.32}                                                                    \\ \cline{3-9} 
                            &                                                                       & 72000                         & 120                       & 0.45                                                                     & 17.50              & 2,099.82          & 1.63                                                                        & 98.94                                                                    \\ \hline
\multirow{7}{*}{Theano}     & \multirow{3}{*}{500}                                                  & 20000                         & 200 (200)                       & \multirow{2}{*}{0.1}                                                     & \DEFAULTHIGHLIGHT{2.80}               & \DEFAULTHIGHLIGHT{560.49}            & \DEFAULTHIGHLIGHT{0.19}                                                                        & \DEFAULTHIGHLIGHT{99.05$\pm$0.00}                                                                    \\ \cline{3-4} \cline{6-9} 
                            &                                                                       & 23000                         & 230 (800)                       &                                                                          & 2.80               & 644.05            & 0.19                                                                        & 99.05                                                                    \\ \cline{3-9} 
                            &                                                                       & 10800                         & 108 (200)                      & 0.25                                                                     & 2.79               & \textbf{301.19}            & 0.19                                                                        & \textbf{99.09}                                                                    \\ \cline{2-9} 
                            & \multirow{4}{*}{5000}                                                 & 8000                         & 800 (800)                      & 0.01                                                                     & 2.51               & 2,006.74          & \textbf{0.18}                                                                        & 97.78                                                                    \\ \cline{3-9} 
                            &                                                                       & 2000                          & 200 (200)                       & \multirow{2}{*}{0.1}                                                     & \textbf{2.50}               & 499.59            & \textbf{0.18}                                                                        & 98.61                                                                    \\ \cline{3-4} \cline{6-9} 
                            &                                                                       & \textbf{8000}                         & \textbf{800 (800)}                      &                                                                          & \textbf{2.50}               & 2,002.6          & \textbf{0.18}                                                                        & \textbf{99.13}                                                                    \\ \cline{3-9} 
                            &                                                                       & 8000                         & 800 (800)                      & 0.25                                                                     & 2.51               & 2,008.56          & \textbf{0.18}                                                                        & \textbf{99.10}                                                                    \\ \hline
\end{tabular}
} 
\vspace{-6mm}
\end{table*}

\subsection{The Impact of Learning Rate}

Above experiments shows that a larger batch size combined with a larger \#Iterations (\#Epochs) may help achieve better accuracy than the default configurations. It also indicates that even with much longer training time, a larger batch size is more resistant to over-fitting. Furthermore, sufficient \#Iterations or \#Epochs are necessary for achieving desirable accuracy, while too few iterations (or epochs) with a larger batch size could lead to under-fitting, hurting the accuracy. We conjecture that seeking a balance between accuracy and training time is more desirable for many DL applications. These observations motivate us to further study the impact of the learning rate (LR). 

Through experiments, we found the series of hyper-parameters that outperform the default w.r.t. accuracy and even runtime performance. Specifically, the batch size, \#Iterations, LR and accuracy found for TensorFlwo, Caffe, Torch and Theano are shown in bold on Table \ref{table:accuracy-improvement-mnist}, which shows the measurement comparison of three hyper-parameters: the batch size, \#Iterations, LR on accuracy of four DL frameworks. Recall Figure \ref{fig:accuracy-improvement-mnist-1}, it shows the accuracy and training time for the four DL frameworks by varying \#Iterations (\#Epochs) under the batch size and LR set as ones in bold respectively on Table \ref{table:accuracy-improvement-mnist} of the corresponding DL framework. We highlight two observations.
First, the training time is proportional to the \#Iterations (\#Epochs). Second, the accuracy tends to increase until it reaches the plateau as the \#Epochs increases.

From Table \ref{table:accuracy-improvement-mnist}, we also observe that by varying the batch size and the LR setting, we can obtain improved accuracy over the corresponding default configuration for each DL framework. Concretely, for TensorFlow, the configuration of $batch\_size = 5000$, $\#Iterations = 60000$ and $LR = 0.001$ achieved the highest accuracy of 99.45\%, compared with 99.21\% of the default setting for TensorFlow with $batch\_size=50$, $\#Iterations=20000$, and the $LR = 0.0001$. This is obtained through a progressive study through measurements: We first changed the LR from 0.0001 (default) to 0.001 to study the impact of LR. Then, we changed the \#Iterations from its default (20000) to 60000. For Caffe, Torch and Theano, we conducted similar experiments. Theano employed an early-stopping mechanism to combat over-fitting~\cite{theano}. During training, Theano will monitor the model performance on a validation dataset (neither the training nor the testing dataset). If the model performance fails to improve sufficiently, or even degrades with further training, the early-stopping mechanism will be triggered to stop the training. Therefore, the values within the parentheses represent the set values that are used by  Theano compared to the actual values in the \#Epochs column. For example, 230 (800) represents the setting of \#Epochs = 800 and Theano stopped training at the 230th epoch by its early-stopping mechanism. From the experimental results in Table \ref{table:accuracy-improvement-mnist}, we highlighted several interesting observations.

(1) Accuracy measures the utility of a trained DNN model. Tuning hyper-parameters can lead to accuracy improvement. TensorFlow, Caffe, Torch and Theano obtained accuracy improvements by 0.24\%, 0.04\%, 0.10\% and 0.07\% respectively. It is widely acknowledged in machine learning (ML) community that even small improvement on accuracy can have significant impact on the utility of the trained DNN model, as demonstrated in recent ML literature for newly proposed DL algorithms~\cite{graph-ngram-learning, he-2015-surpass-human-level-performance-imagenet, mishkin2015all}. For instance, the accuracy over the conventional algorithms is improved by 0.02\% on CVPR 2012~\cite{cirecsan2012multi} and on ICML 2013~\cite{wan2013regularization}, and such small percentage is non-trivial when it is over a large dataset of 100,000 in size.


(2) Accuracy is sensitive to the setting of LR while different settings of LR show little impact on the training time and testing time. For all four DL frameworks, slight changes in LR led to significant accuracy variance. For TensorFlow, when the LR is changed from 0.001 to 0.01, the accuracy of its trained DNN dropped from 99.45\% to 98.59\%. For Caffe, when the LR is changed from 0.1 to 0.01, the accuracy of its trained DNN dropped by 0.18\%, from 99.18\% to 99.00\%. It is also worth to note that for the LR=0.1 with batch\_size=64 and LR=0.5 with batch\_size=6400, Caffe trained DNN failed to converge, denoted by NaN for accuracy due to the improper setting~\cite{GT-ICDCS2018paper}. Table \ref{table:accuracy-improvement-mnist} also shows that the training time and testing time are kept almost the same for each of the four DL frameworks when we only vary the settings of LR. For example, the training time of TensorFlow with batch\_size=5000 and \#Iterations=60000 for LR=0.0001, 0.001 and 0.01 are 8484.47s, 8379.51s and 8331.56s respectively, showing a small variance. This observation also indicates that tuning LR may lead to higher accuracy with negligible runtime performance degradation.

(3) The training time per epoch is highly dependent on the batch size. Also the total training time is proportional to the \#Iterations under a fixed batch size. All four frameworks manifest similar training time per epoch for a specific batch size. For example, the training time per epoch for Caffe with batch\_size=64 is 9.10s$\sim$9.16s while it is 7.66s $\sim$ 7.72s with batch\_size=6400. Similar observations are found in other frameworks, indicating that the training time is somewhat more predictable for DL frameworks, namely, under a fixed batch size, the total training time is proportional to the \#Iterations (\#Epochs).

(4) The impact of combined hyper-parameters is independent of the optimal settings of individual hyper-parameters. For example, the default configuration of TensorFlow can achieve higher accuracy with either a larger \#Iterations (60000, 99.38\%) or a larger LR (0.001, 99.26\%). However, when we combine the larger \#Iterations (60000) and LR (0.001) and compare the combination to the default setting, the accuracy dropped from 99.21\% to 99.11\%. This also indicates that the complexity of finding the optimal settings for multiple hyper-parameters, which is another reason that makes the tuning and benchmarking of DL framework a more challenging compared to conventional big data processing systems. 


(5) In addition to improving the accuracy of default hyper-parameters, we observe that two sets of hyper-parameters in Torch and Theano outperform the default ones on both accuracy and training time performance. Specifically, for Torch, batch\_size=100, \#Epochs=12 and LR=0.4 surpassed its default configuration by a shorter training time of 209.64 seconds compared to 338.46 seconds, and a higher accuracy of 99.24\% compared to 99.22\% for the Torch default setting. For Theano, the combination of batch\_size=500, \#Epochs=200 and LR=0.25 outperforms the default configuration by 301.19s over 560.49s for training time and 99.09\%  over 99.06\% for accuracy. Notably, these two setting combinations also reduced the training time to approximately 60\% of the training time when using their default settings.


\subsection{Impact of CPU and Memory Resource} 
\label{section:cpu-memory}

We have shown the feasibility of tuning individual hyper-parameters and tuning multiple hyper-parameters for improving the accuracy of the DL frameworks over their default configurations. In this section, we examine how different DL frameworks respond to different batch sizes with respect to their CPU and memory resource usage patterns, given that larger batch sizes may demand more CPU processing and consume more memory resource. Table~\ref{table:cpu-memory-batchsize-cpu-1} shows the CPU and memory usage measurement results with varying batch sizes on MNIST for all four frameworks.



\begin{table}[h!]
\vspace{-3mm}
\centering
\caption{CPU/Memory Usage by Varying Batch Size on MNIST (CPU)}
\label{table:cpu-memory-batchsize-cpu-1}
\vspace{-4mm}
\scalebox{0.75}{
\small
\begin{tabular}{|c|c|c|c|c|c|c|c|}
\hline
\multirow{2}{*}{Framework} & \multirow{2}{*}{\begin{tabular}[c]{@{}c@{}}Batch\\ Size\end{tabular}} & \multicolumn{4}{c|}{CPU Usage (\% AVG)} & \multicolumn{2}{c|}{Memory (MB)} \\ \cline{3-8} 
                           &                                                                       & user    & system   & iowait   & total   & AVG             & MAX            \\ \hline
\multirow{4}{*}{TF}        & 50                                                                    & 72.26   & 7.88     & 0.01     & 80.15   & 736.73          & 6,268.55       \\ \cline{2-8} 
                           & 500                                                                   & 80.30   & 6.30     & 0.01     & 86.61   & 764.61          & 6,204.89       \\ \cline{2-8} 
                           & 5000                                                                  & 81.91   & 6.90     & 0.01     & 88.82   & 1,812.38        & 6,282.23       \\ \cline{2-8} 
                           & 50000                                                                 & 65.31   & 5.60     & 7.65     & 78.56   & 11,203.91       & 31,724.76      \\ \hline
\multirow{4}{*}{Caffe}     & 64                                                                    & 31.41   & 17.89    & 0.01     & 49.31   & 462.92          & 484.38         \\ \cline{2-8} 
                           & 640                                                                   & 31.11   & 18.22    & 0.01     & 49.34   & 559.03          & 584.04         \\ \cline{2-8} 
                           & 6400                                                                  & 30.07   & 14.71    & 0.02     & 44.80   & 1,546.65        & 1,597.41       \\ \cline{2-8} 
                           & 60000                                                                 & 29.58   & 11.26    & 0.02     & 40.86   & 10,395.47       & 11,009.01      \\ \hline
\multirow{4}{*}{Torch}     & 10                                                                    & 34.64   & 65.20    & 0.00     & 99.84   & 661.51          & 768.32         \\ \cline{2-8} 
                           & 100                                                                   & 37.94   & 58.76    & 0.00     & 96.70   & 572.37          & 622.91         \\ \cline{2-8} 
                           & 1000                                                                  & 32.29   & 37.26    & 0.01     & 69.56   & 1,146.22        & 1,340.66       \\ \cline{2-8} 
                           & 10000                                                                 & 37.48   & 42.61    & 0.00     & 80.09   & 8,929.00        & 9,110.44       \\ \hline
\multirow{3}{*}{Theano}    & 50                                                                    & 48.32   & 51.59    & 0.00     & 99.91   & 773.00          & 885.63         \\ \cline{2-8} 
                           & 500                                                                   & 41.16   & 44.33    & 0.00     & 85.49   & 1,621.72        & 4,062.09       \\ \cline{2-8} 
                           & 5000                                                                  & 37.01   & 22.85    & 0.01     & 59.87   & 1,458.26        & 2,349.91       \\ \hline
\end{tabular}
} 
\vspace{-3mm}
\end{table}

\subsubsection{TensorFlow}
We make two interesting observations on the CPU usage of TF. First, as the batch size increases from 50 to 500 and 5000, the CPU usage of TF (total) increases accordingly with almost no \%iowait, because the corresponding maximum memory usage for all three batch size settings is within the total of the 32GB memory of Server 1. However, when the batch size is increased to 50000, the percentage of the CPU usage for user mode (\%user) drops significantly, the \%iowait increases, and the maximum memory consumed is slightly over 31GB, very close to the physical memory capacity. The increased \%iowait shows a heavy disk read/write during the execution, indicating that memory swapping occurs, which degrades the overall system performance with much longer training time. This also implies that the adequate batch size with respect to the physical memory capacity is critical for ensuring high performance of DL frameworks.

\subsubsection{Caffe}
From Table~\ref{table:cpu-memory-batchsize-cpu-1}, we measure the performance of Caffe by varying its batch size from the default value of 64 to 640, 6400 and 60000, we observe that the total CPU usage is decreasing as the batch size increases. However, the average memory consumption and the maximum memory consumption show different responses as the batch size increases to 640, 6400 and 60000. For example, the maximum memory for Caffe is 11009.01MB, about 11GB, when the batch size is increased to 60000. Compared with TensorFlow, Caffe uses much less memory to keep the whole batch of the training dataset in the memory, while TensorFlow runs out of the memory for $batch\_size = 50000$, possibly due to the data-flow centric processing method of TensorFlow, which introduces more intermediate data structures. In comparison, for the default batch size of 64, the maximum memory usage is about 484.38 MB. This is sufficient to accommodate the in-memory data size of 209.62 MB (179.67 MB+29.95 MB) for MNIST~\cite{mnistlenet}. When the batch size is increased to 60000, the in-framework data expands to 50.21$\times$ ((11009.01-484.38)/209.62), compounded with the fact that the feature map of the training dataset may occupy a huge amount of memory~\cite{tbd},
making the execution of Caffe memory-intensive. One take-away from this empirical analysis is the potential of improving the CPU usage for large batch sizes, as the higher CPU usage accounts for faster training with shorter time to training completion.

\subsubsection{Torch \& Theano}
Similarly, the CPU usage drops as the batch size increases for Torch and Theano, even though the memory usage increases significantly for Torch when batch size changes from 10~to~10000 and for Theano when the $batch\_size$ changes from 50~to~500.

Intuitively, the larger batch size will increase the workload for parallel computations, which should consume more CPU, and thus the CPU usage is expected to increase. However, we observe from Table~\ref{table:cpu-memory-batchsize-cpu-1} that the CPU usage drops as the batch size increases for all four DL frameworks. This further indicates that the CPU usage is not efficient for larger batch sizes, and optimizations that can further improve CPU usage may further speed up the training process. Moreover, our experiments also show that a large batch size increases the memory usage and reduces the training time, demonstrating the feasibility of space and time tradeoff. These observations further indicate that improving the CPU and memory usage hold the potential to further optimize the performance.

\section{Related Work and Conclusion}
We have presented a systematic approach to empirical analysis and characterization of four popular DL frameworks, TensorFlow, Caffe, Torch and Theano, on three representative datasets, MNIST, CIFAR-10 and ImageNet. This paper makes three unique contributions. First, some existing benchmarking efforts for DL frameworks \cite{convnet-benchmarks, deepbench} suffer from two inherent problems: (1) they measure the average time for forward and backward passes, matrix multiplication, or layer-wise performance, and lack of overall performance characterization; and (2) they do not include accuracy comparison. Although some recent efforts~\cite{dawnbench, tbd} provide end-to-end DL benchmarking by only focusing on the training phase or specific DL tasks, none to date has taken a holistic approach to study the impact of hardware configurations, parallel computing libraries, and hyper-parameters on the performance of DL frameworks with respect to both accuracy and training time. Second, to the best of our knowledge, this study is the first to identify the opportunities for configuring parallel computing libraries and tuning individual and multiple hyper-parameters for improving the training time performance and the accuracy of DL frameworks. Third but not the least, to gain a deeper understanding of the impact of hyper-parameters and the choice of parallel computing libraries on the accuracy and training time of DL frameworks, we provide a systematic analysis of the CPU and memory usage patterns for different parallel computing libraries and different batch sizes and their impact on accuracy and training efficiency. 
We conjecture that the comparative measurement study and analysis  presented in this paper will provide empirical guidance for developers, service providers to provide high performance DLaaS with better performance and accuracy tuning tools and at the same time it also helps application developers and end-users to select the right DL frameworks for the right DL workloads.

\subsection*{Acknowledgment}
This research is partially sponsored by National Science Foundation under CISE SAVI/RCN (1402266, 1550379), CNS (1421561), CRISP (1541074), SaTC (1564097) programs, an REU supplement (1545173), an IBM Faculty Award, and gifts, grants, or contracts from Fujitsu, HP, Intel, and Georgia Tech Foundation through the John P. Imlay, Jr. Chair endowment. 

\ifCLASSOPTIONcaptionsoff
  \newpage
\fi



\bibliographystyle{IEEEtran}
\bibliography{reference}
%



%


\vspace{-15mm}
\begin{IEEEbiography}[{\includegraphics[width=1in,height=1.25in,clip,keepaspectratio]{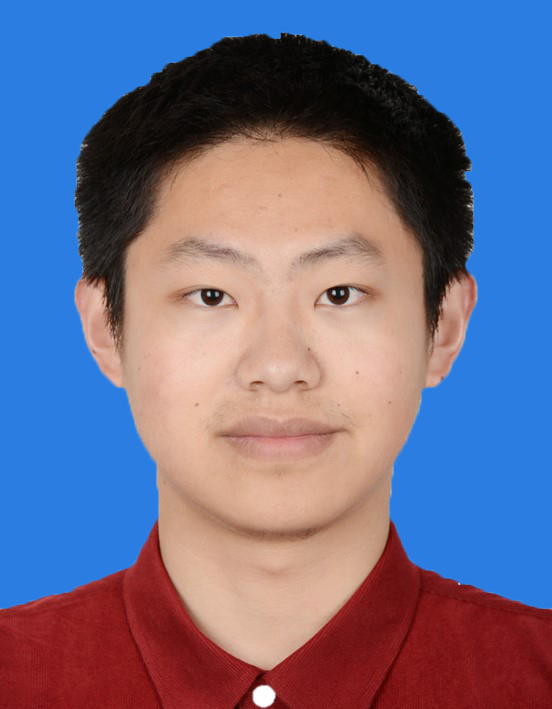}}]{Yanzhao Wu} is currently a PhD student in the School of Computer Science, Georgia Institute of Technology. He received his B.E. degree from the School of Computer Science and Technology, University of Science and Technology of China. His research interests are centered primarily on systems for machine learning and big data, and machine learning algorithms for optimizing systems.
\end{IEEEbiography}
\vspace{-15mm}
\begin{IEEEbiography}[{\includegraphics[width=1in,height=1.25in,clip,keepaspectratio]{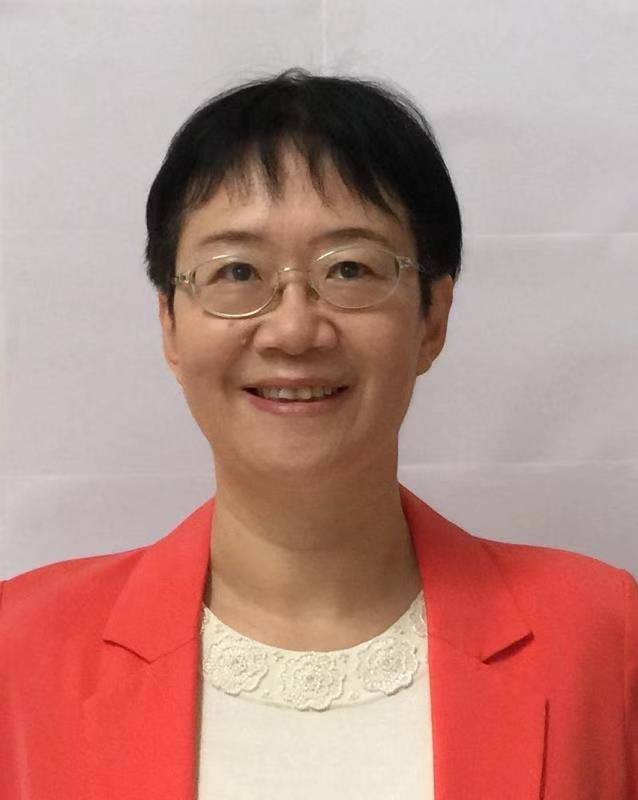}}]{Ling Liu} is a professor in the School of Computer Science, Georgia Institute of Technology, and directs the research programs in the Distributed Data Intensive Systems Lab (DiSL). She is an elected IEEE fellow, a recipient of the IEEE Computer Society Technical Achievement Award in 2012, and a recipient of the best paper award from a dozen of top venues, including ICDCS, WWW, Pat Goldberg Memorial Best Paper Award, IEEE Cloud, IEEE ICWS, ACM/IEEE CCGrid, IEEE Symposium on Big Data, IEEE IoT, IEEE Edge. She is the Editor in Chief of ACM Transactions on Internet Computing (Dec. 2018 onward). Her current research is primarily sponsored by NSF and IBM.
\end{IEEEbiography}
\vspace{-15mm}
\begin{IEEEbiography}[{\includegraphics[width=1in,height=1.25in,clip,keepaspectratio]{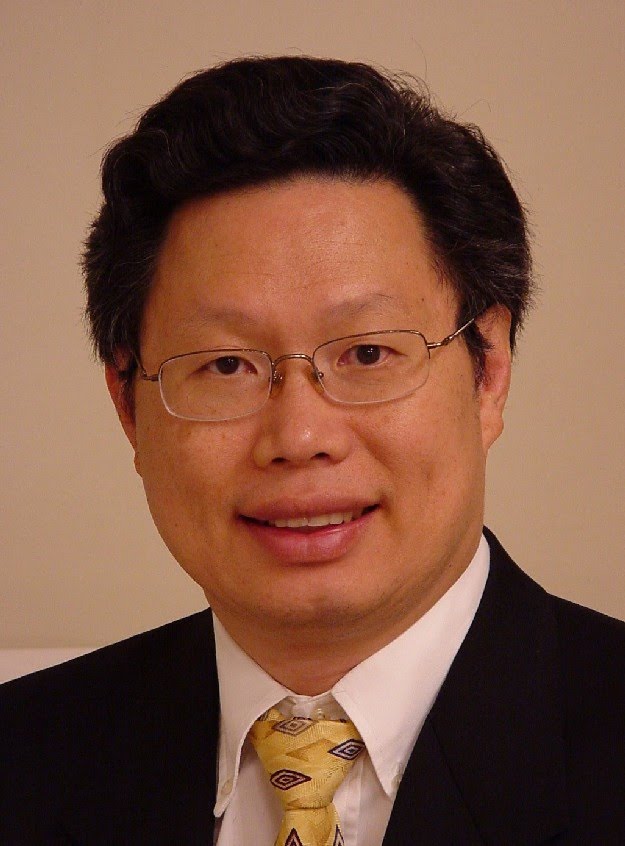}}]{Calton Pu}
is a professor and holds John P. Imlay, Jr. Chair in Software in the College of Computing, Georgia Institute of Technology. He has worked on several projects in systems and database research. He has published more than 70 journal papers and book chapters, 270 conference and refereed workshop papers. He served on more than 120 program committees, more than 20 times as general (co-)chair and PC (co-)chairs. His recent research has focused on big data in Internet of Things, automated N-tier application performance, and denial of information. He is a fellow of AAAS and IEEE.
\end{IEEEbiography}
\vspace{-15mm}
\begin{IEEEbiography}[{\includegraphics[width=1in,height=1.25in,clip,keepaspectratio]{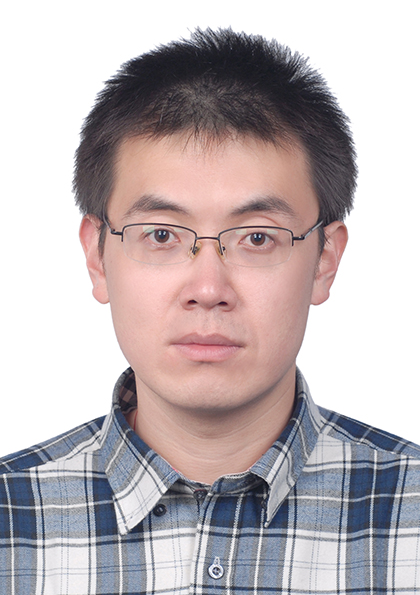}}]{Wenqi Cao} is currently pursuing Computer Science PhD degree and is a research assistant in the School of Computer Science at Georgia Institute of Technology. Prior to that, he received his Master Degree from the Computer Science and Engineering Department at Pennsylvania State University, and has five-years work experience as software engineer for SIEMENS, HP and Microsoft. His research interest includes cloud computing, distributed system, operating system, in-memory system, programming language, machine learning, performance evaluation, algorithm and information retrieval.
\end{IEEEbiography}
\vspace{-15mm}
\begin{IEEEbiography}[{\includegraphics[width=1in,height=1.25in,clip,keepaspectratio]{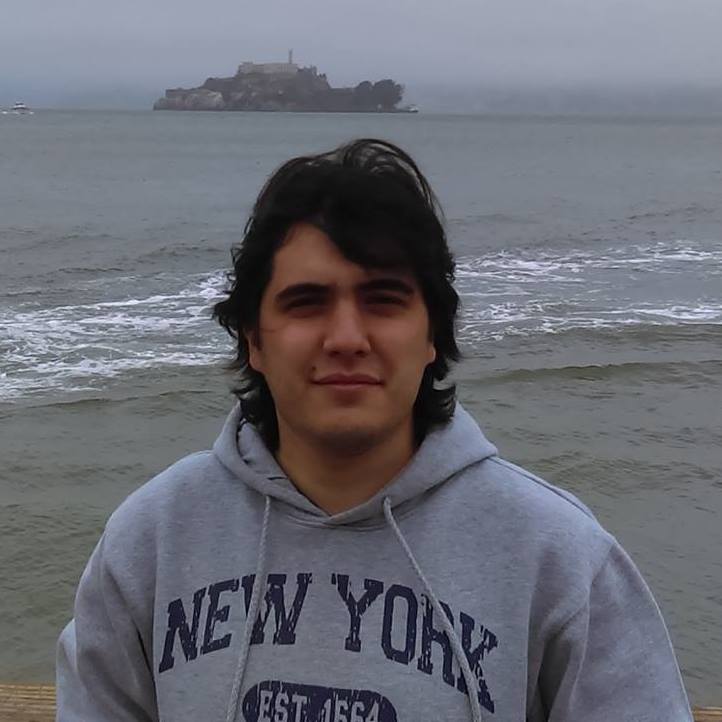}}]{Semih Sahin} is currently a Ph.D. candidate at College of Computing, Georgia Tech, USA. He obtained M.S degree in Computer Engineering from Bilkent University, Turkey. His research interests are in large scale data processing and machine learning, including stream processing and memory management in big data frameworks.
\end{IEEEbiography}
\vspace{-150mm}
\begin{IEEEbiography}[{\includegraphics[width=1in,height=1.25in,clip,keepaspectratio]{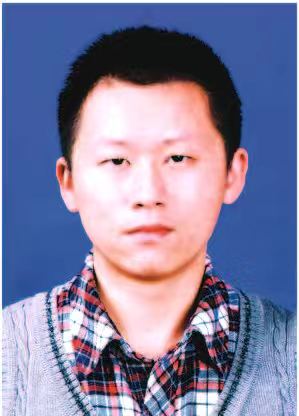}}]{Wenqi Wei} is currently pursuing his PhD in the School of Computer Science, Georgia Institute of Technology. He received his B.E. degree from the School of Electronic Information and Communications, Huazhong University of Science and Technology. His research interests include data privacy, machine learning security and trust, innovative machine learning algorithms for big data systems and services.
\end{IEEEbiography}
\vspace{-150mm}
\begin{IEEEbiography}[{\includegraphics[width=1in,height=1.25in,clip,keepaspectratio]{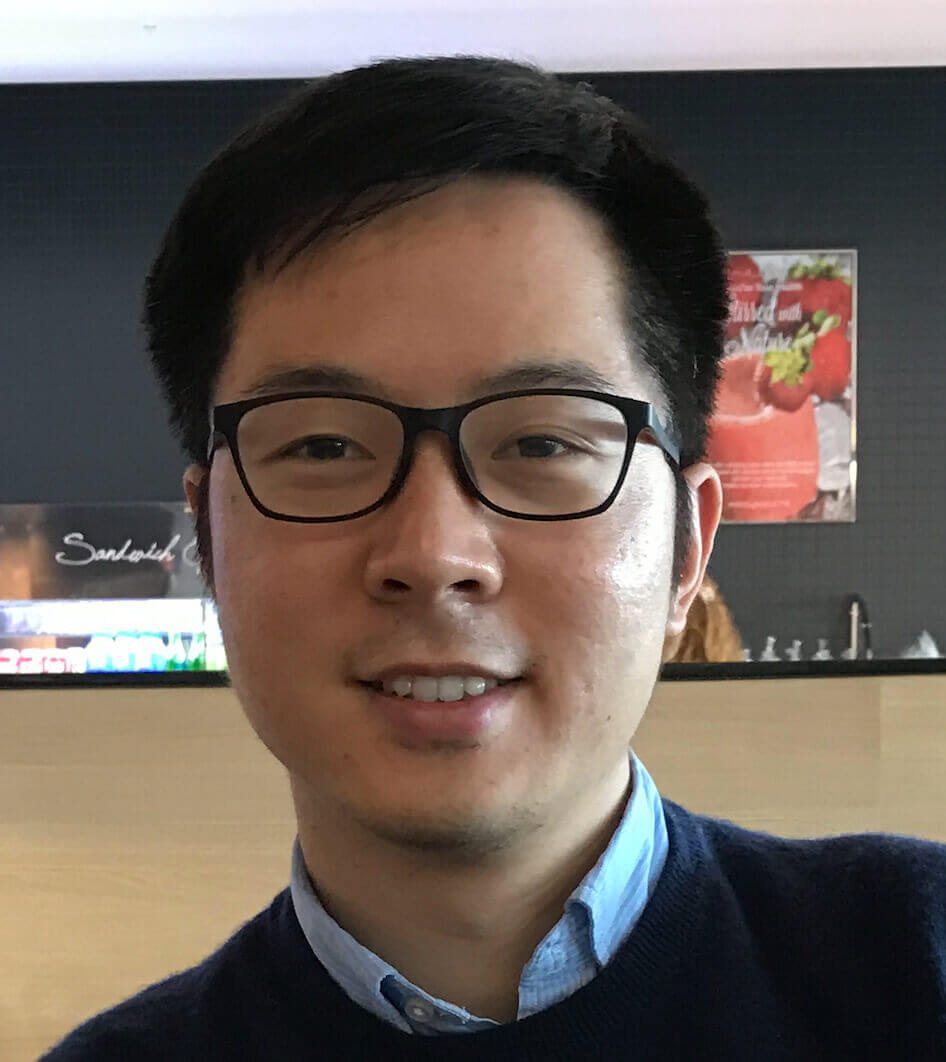}}]{Qi Zhang} is a Research Staff Member in IBM Thomas J. Watson Research Center, received the Ph.D. degree in computer science from Georgia Institute of Technology (USA) in 2017. His research interests include blockchain systems, cloud computing, big data and deep learning systems, and published in major referred journals, such as IEEE TC, IEEE TSC, ACM CSUR, IEEE Blockchain NewsLetter, and conferences, such as VLDB, Blockchain, IEEE ICDCS, SuperComputing, ICWS, CLOUD, and VLDB, HPDC.
\end{IEEEbiography}







\end{document}